\documentclass[11pt,a4paper,oneside,english]{article}
\usepackage{babel}
\usepackage[utf8]{inputenc}
\usepackage[T1]{fontenc} 
\usepackage{amsmath}
\usepackage{amssymb}
\usepackage{physics}
\usepackage{graphicx}
\usepackage{booktabs}
\usepackage[hidelinks]{hyperref}
\usepackage{makecell}
\usepackage{romannum}
\usepackage{longtable}
\usepackage[flushleft]{threeparttable}
\usepackage{cite}
\usepackage{fullpage,authblk}
\usepackage{float}
\usepackage{blindtext}
\usepackage{lmodern}
\usepackage[numbers]{natbib}

\providecommand{\keywords}[1]{
  \small	
  \textbf{\textit{Keywords---}} #1
}

\begin{document}

\title{\textbf{Efficient Mediated Semi-Quantum Key Distribution Protocol Using Single Qubits}}

\author[1]{Mustapha Anis Younes \footnote{Corresponding Author: \texttt{mustaphaanis.younes@univ-bejaia.dz}}}
\author[2]{Sofia Zebboudj \footnote{\texttt{sofiazebboudj@gmail.com}}}
\author[3]{Abdelhakim Gharbi \footnote{\texttt{abdelhakim.gharbi@univ-bejaia.dz}}}
\affil[1,2]{Université de Bejaia, Faculté des Sciences Exactes, Laboratoire de Physique Théorique, 06000 Bejaia, Algérie}
\affil[2]{ENSIBS, Université Bretagne Sud, 56000 Vannes, France}
\date{\vspace{-8ex}}

\maketitle 
\pagenumbering{arabic}

\begin{abstract}
    In this paper, we propose a new efficient mediated semi-quantum key distribution (MSQKD) protocol, facilitating the establishment of a shared secret key between two classical participants with the assistance of an untrusted third party (TP). Unlike existing MSQKD protocols, our approach significantly reduces the quantum requirements for TP, who only needs to prepare and measure qubits in the $X$ basis. Meanwhile, the classical participants are limited to preparing and measuring qubits in the $Z$ basis, along with performing Hadamard operations. This reduction in quantum overhead enhances the practicality of our MSQKD protocol without compromising qubit efficiency. Additionally, we demonstrate the security of our protocol against various well-known attacks.
   
\end{abstract}

 \keywords{Quantum cryptography, Semi-quantum key distribution, Mediated key distribution protocol, Dishonest third party, Single qubits.}

\section{Introduction}\label{sec1}

Quantum key distribution (QKD), pioneered by Bennett and Brassard in 1984\cite{Bennett1992}, enables two participants to securely share a secret key by exploiting the principles of quantum mechanics. Since then, numerous QKD protocols have been proposed \cite{Long2002, ZHANG2005,Li2008,Shu2023,Sharma2021,Lin2012,Lo2005,Lo2012,Hwang2003,Lai2020,Shor2000}. However, all those protocols are designed under the premise that all
participants in quantum communication possess full quantum capabilities. This assumption, however, is highly unrealistic given the high costs and complexity involved in the devices. Challenges such as quantum state preparation, storage, and transmission only add to this complexity \cite{Cirac2020}. Therefore, it's not practical to expect all participants to afford such expensive devices and carry out these intricate operations.

In response, Boyer et al. \cite{Boyer2007} introduced the concept ”semi-quantum
cryptography” in 2007. This approach allows one party to possess full quantum capabilities while limiting the quantum capacity of the other, facilitating secure communication between a quantum and a classical user. Specifically, a participant capable of: (a) generating any quantum states, (b) conducting any quantum measurements, and (c) storing qubits in quantum memory, is termed a quantum participant. Conversely, a classical participant is one with the ability to perform three of the following operations: (1) generating qubits with the $Z$ basis $\{\ket{0}, \ket{1}\}$; (2) measuring qubits with the $Z$ basis $\{\ket{0}, \ket{1}\}$; (3) reflecting qubits without disturbance; and (4) reordering qubits.

Since the introduction of the first semi-quantum protocol, many other semi-quantum protocols have emerged, including SQKD \cite{Boyer2009,Zou2009,Wang2011,Yu2014,Li2016,Zou2015,He2018,Li2016a,ZHIWEI2013,Yu2017,Boyer2017,Zhou2019,Tsai2020}, semi-quantum communications \cite{Zou2014,Luo2015,Shukla2017}, semi-quantum secret sharing protocols (SQSS) \cite{Xiang2019,Li2010,Xie2015}, and semi-quantum private comparisons (SQPC) \cite{Lin2019a,Chongqiang2021,Chou2016}. Additionally, in 2013, Nie et al. \cite{Nie2012} introduced unitary operations on a single qubit as a viable operation in a semi-quantum environment, expanding beyond Boyer et al.'s initial set of operations. Subsequently, some semi-quantum protocols \cite{Tsai2020,Tsai2021,Li2020,Tsai2022,Tsai2019,Yang2020} have incorporated unitary operations within the range of quantum operations permissible for classical participants, in addition to Boyer et al.'s original four operations. This inclusion of unitary operations in semi-quantum environments was mainly motivated by the rapid development and feasibility of technology related to unitary operations for single qubits \cite{Iqbal2020,Bouwmeester1997,Ren2017,Young2013,Tipsmark2011,Cerf1998,O’Brien2007}, enabling real-world implementation. For an in-depth review of semi-quantum cryptography, readers are directed to \cite{Iqbal2020}, which provides a comprehensive survey on the subject.

Although Semi-Quantum Key Distribution (SQKD) protocols offer practical advantages over traditional Quantum Key Distribution (QKD) methods, only the user with full quantum capabilities can share a secret. While this aligns with practical considerations, it prompts an interest in exploring how a classical user could share their secret and remain protected from a dishonest quantum user. This was first investigated by Krawec \cite{Krawec2015} in 2015, who introduced the first mediated semi-quantum key distribution (MSQKD) protocol. This protocol enables two "classical" participants to establish a shared secret key securely with the help of an untrusted quantum third party (TP), who may attempt any possible attack. Krawec protocol's is based on Bell states, with classical participants limited to the quantum capabilities previously outlined as (1), (2), and (3). In the same year, Krawec \cite{Krawec2015a} also proposed an enhanced asymptotic key rate bound for the protocol. Several other mediated protocols have since been introduced,  aiming to reduce the quantum requirements of the untrusted TP and enhance overall efficiency.

 In 2018, Liu et al. \cite{Liu2018} proposed a protocol based on entanglement swapping of Bell states, relieving classical users of the need for quantum measurement capability. However, vulnerabilities were later identified by Zou et al. \cite{Zou2020}. In 2019, Lin et al. \cite{Lin2019} explored reducing the quantum overhead of the TP by utilizing single photons as resource states, though with limited efficiency gains. Massa et al. \cite{Massa2019} recently introduced a novel MSQKD protocol where classical participants require access only to superimposed single photons as a viable quantum resource, successfully demonstrating it in experiments.

Moreover, in 2019, Tsai et al. \cite{Tsai2019} proposed a lightweight MSQKD protocol capable of mitigating Trojan horse attacks through one-way quantum communication. Nonetheless, this protocol assumes a trustworthy TP, which may not always be practical. In the same year, Tsai and Yang \cite{Tsai2021} introduced another lightweight MSQKD protocol, also resilient against Trojan horse attacks with a dishonest TP, based on Bell states and requiring classical participants to: (1) perform a measurement in the $Z$ basis and (2) perform the Hadamard operation.

Further advancements continued in subsequent years. Yang and Hwang, \cite{Yang2020} in 2020, introduced an MSQKD protocol accommodating classical participants with limited asymmetric quantum capabilities. In 2021, Chen et al. \cite{Chen2021} proposed an MSQKD protocol employing single qubits in both the initial preparation stage and subsequent server measurement, demonstrating improved efficiency compared to previous approaches. In 2022, Guskind and Krawec \cite{Guskind2022} introduced a new MSQKD boasting asymptotically perfect efficiency, albeit with reduced noise tolerance. Finally, in 2023, Ye et al. \cite{Ye2022} introduced a circular MSQKD protocol based on Bell states, establishing its unconditional security in the asymptotic scenario.

In this paper, we propose a new MSQKD protocol based on single qubits with the help of an untrusted third party (TP). Our protocol demonstrates improved efficiency compared to similar approaches and notably reduces the quantum overhead of the TP, requiring only two key operations: (a) generate single qubits in the state $\ket{+}=\frac{1}{\sqrt{2}}(\ket{0}+\ket{1})$, (b) measure qubits in the $X$ basis $\{\ket{+}=\frac{1}{\sqrt{2}}(\ket{0}+\ket{1}), \ket{-}=\frac{1}{\sqrt{2}}(\ket{0}-\ket{1})\}$, thereby enhancing its practicality.  As for the classical participants, they are restricted to performing the following operations: (1) generate qubits in the $Z$ basis $\{\ket{0}, \ket{1}\}$, (2) measure qubits in the $Z$ basis, and (3) perform the Hadamard operation on individual qubits. Furthermore, we prove that our protocol is secure against measurement attack, Faked states attack, and collective attack in the ideal situation.

The subsequent sections are structured as follows: \hyperref[sec2]{Section 2} introduces our MSQKD protocol, while \hyperref[sec3]{Section 3} delves into its security analysis in ideal scenarios. \hyperref[sec4]{Section 4} offers a comparative analysis with similar MSQKD protocols, highlighting the efficiency and practicality of our approach. Finally, the paper concludes in the \hyperref[sec5]{last section}.

\section{The proposed protocol}\label{sec2}

In this section, we present our mediated SQKD protocol in a step-by-step manner. The scenario involves two classical participants, Alice and Bob, aiming to establish a shared secret key with the assistance of an untrusted third party (TP). The TP is only required to prepare and measure qubits in the $X$ basis $\{\ket{+}, \ket{-}\}$, where $\ket{+}=\frac{1}{\sqrt{2}}(\ket{0}+\ket{1})$, and $\ket{-}=\frac{1}{\sqrt{2}}(\ket{0}-\ket{1})$. A qubit measurement in the $X$ basis is defined by the $X$ measurement operators:
\begin{equation}\label{eq1}
    \begin{aligned}
        M_{+}&=\ket{+}\bra{+}=\frac{1}{2}(\ket{0}+\ket{1})(\bra{0}+\bra{1})=\frac{1}{2}\begin{pmatrix}
            1 & 1 \\
            1 & 1
        \end{pmatrix},  \\
        M_{-}&=\ket{-}\bra{-}=\frac{1}{2}(\ket{0}-\ket{1})(\bra{0}-\bra{1})=\frac{1}{2}\begin{pmatrix}
            1 & -1 \\
            -1 & 1
        \end{pmatrix}.
    \end{aligned}
\end{equation}

On the other hand, Alice and Bob are limited to executing the following operations:

\begin{enumerate}
    \item Generating qubits in the $Z$ basis $\{\ket{0}, \ket{1}\}$.
    \item Measuring qubits in the $Z$ basis, such that the measurement of a qubit in the $Z$ basis is defined by the $Z$ measurement operators:
    \begin{equation}\label{eq2}
    \begin{aligned}
        M_{0}&=\ket{0}\bra{0}=\begin{pmatrix}
            1 & 0 \\
            0 & 0
        \end{pmatrix},  \\
        M_{1}&=\ket{1}\bra{1}=\begin{pmatrix}
            0 & 0 \\
            0 & 1
        \end{pmatrix}.
    \end{aligned}
\end{equation}
    \item Performing the Hadamard operation, which can be defined as follow:
    \begin{equation}\label{eq3}
        H=\frac{1}{\sqrt{2}}(\ket{0}\bra{0}+\ket{1}\bra{0}+\ket{0}\bra{1}-\ket{1}\bra{1})=\frac{1}{\sqrt{2}}\begin{pmatrix}
            1 & 1 \\
            1 & -1
        \end{pmatrix}.
    \end{equation}
\end{enumerate}

The quantum channels between the participants are assumed to be ideal (i.e. non-lousy and noise-free). Additionally, there exist public classical channels between each pair of participants, with the communication channel between Alice and Bob being authenticated. The procedure of our proposed MSQKD protocol unfolds as follows:\\

\textbf{Step 1:} TP generates $2N$ qubits and sends them individually to Alice. Note that each qubit is initially prepared in the state $\ket{+}$. \\

\textbf{Step 2:} Upon receiving a qubit from the TP, Alice randomly selects one of the two operations: 
\begin{itemize}
    \item \textbf{MH \text{(Measure then Hadamard)}:} This operation entails measuring the received qubit in the $Z$-basis $\{\ket{0}, \ket{1}\}$, and generating a new qubit in the same state as the measured one, then applying a Hadamard operation on this newly generated qubit before passing it on to Bob. When Alice chooses this operation, she measures the state $\ket{0}$ or $\ket{1}$ with equal probability:
    \begin{equation}\label{eq4}
        \begin{aligned}
            P(0)&= \bra{+}M_{0}\ket{+} = \abs{\braket{0}{+}}^{2}=\frac{1}{2}, \\
            P(1)&= \bra{+}M_{1}\ket{+} = \abs{\braket{1}{+}}^{2}=\frac{1}{2}.
        \end{aligned}
    \end{equation}
    This implies that Alice sends the state $\ket{+}$ or $\ket{-}$ with equal probability to Bob since:
    \begin{equation}\label{eq5}
        \begin{aligned}
            \ket{+}&= H\ket{0} = \frac{1}{\sqrt{2}}(\ket{0}\bra{0}+\ket{1}\bra{0}+\ket{0}\bra{1}-\ket{1}\bra{1})\ket{0}, \\
            \ket{-}&= H\ket{1} = \frac{1}{\sqrt{2}}(\ket{0}\bra{0}+\ket{1}\bra{0}+\ket{0}\bra{1}-\ket{1}\bra{1})\ket{1}.
        \end{aligned}
    \end{equation}
    \item \textbf{HM \text{(Hadamard then Measure)}:} Here, the received qubit undergoes a Hadamard operation, followed by a measurement in the $Z$ basis. Subsequently, a new qubit is generated in the same state as the measured one and forwarded to Bob. When Alice selects this operation, after performing the Hadamard operation:
    \begin{equation}\label{eq6}
        \ket{0}= H\ket{+} = \frac{1}{2}(\ket{0}\bra{0}+\ket{1}\bra{0}+\ket{0}\bra{1}-\ket{1}\bra{1})(\ket{0}+\ket{1}),
    \end{equation}
    she will measure the state $\ket{0}$ with certainty since:
    \begin{equation}\label{eq7}
        P(0)=\bra{0}M_{0}\ket{0} = \abs{\braket{0}{0}}^{2} = 1.
    \end{equation}
\end{itemize} 
Thus, the states that Alice can transmit to Bob are $\ket{+}$, $\ket{-}$ with a probability of $\frac{1}{4}$ each, and $\ket{0}$ with a probability of $\frac{1}{2}$. \\

\textbf{Step 3:} After receiving a qubit from Alice, Bob performs the same operations as Alice did in step 2 but sends the resulting qubit to TP. When Bob chooses to perform the \textbf{MH} operation, after his measurement in the $Z$ basis, he will obtain the state $\ket{0}$ or $\ket{1}$ with the following probability: 
\begin{equation}\label{eq8}
    \begin{aligned}
        P(0)&= \bra{+}M_{0}\ket{+} = \abs{\braket{0}{+}}^{2}= \bra{-}M_{0}\ket{-} = \abs{\braket{0}{-}}^{2}=\frac{1}{2}, \\
        P(1)&= \bra{+}M_{1}\ket{+} = \abs{\braket{1}{+}}^{2}= \bra{-}M_{1}\ket{-} = \abs{\braket{1}{-}}^{2}=\frac{1}{2},
    \end{aligned}
\end{equation}
if he receives $\ket{+}$ or $\ket{-}$, and 
\begin{equation}\label{eq9}
\begin{aligned}
    P(0)&=\bra{0}M_{0}\ket{0} = \abs{\braket{0}{0}}^{2} = 1, \\
    P(1)&=\bra{0}M_{1}\ket{0} = \abs{\braket{1}{0}}^{2} = 0. 
\end{aligned}
\end{equation}
if he receives $\ket{0}$. Thus, after performing Hadamard, Bob will either transmit $\ket{+}=H\ket{0}$ or $\ket{-}=H\ket{1}$ depending on his measurement result. On the other hand, when Bob chooses to perform the \textbf{HM} operation, after initially applying the Hadamard operation, depending on the state sent by Alice, he will end up with one of the following states:
\begin{equation}\label{eq10}
    \ket{0}=H\ket{+}, \quad \ket{1}=H\ket{-}, \quad \ket{+}=H\ket{0}.
\end{equation}
After performing his measurement in the $Z$ basis, he will transmit $\ket{0}$ or $\ket{1}$ depending on his measurement outcome. Note that in both step 2 and step 3, Alice and Bob register the measurement outcome as 0 if they obtain the state $\ket{0}$, and as 1 otherwise.
\\

\textbf{Step 4:} Upon receiving a qubit from Bob, the TP conducts a measurement in the $X$-basis $\{\ket{+}, \ket{-}\}$ on the qubit. We know that the states that TP can receive are $\{\ket{0}, \ket{1}, \ket{+}, \ket{-}\}$ with a certain probability for each as displayed in \hyperref[table 1]{Table 1}. When the TP receives a state expressed in the $X$ basis, he will measure that state with absolute certainty:
\begin{equation}\label{eq11}
    \begin{aligned}
        P(+)&= \bra{+}M_{+}\ket{+} = \abs{\braket{+}{+}}^{2} = 1, \\
        P(-)&= \bra{-}M_{-}\ket{-} = \abs{\braket{-}{-}}^{2} = 1.
    \end{aligned}
\end{equation}
On the other hand, if the TP receives a state expressed in the $Z$ basis, he will obtain either $\ket{+}$ or $\ket{-}$ with equal probability:
\begin{equation}\label{eq12}
    \begin{aligned}
        P(+)&= \bra{0}M_{+}\ket{0} = \abs{\braket{+}{0}}^{2}= \bra{1}M_{+}\ket{1} = \abs{\braket{+}{1}}^{2}=\frac{1}{2}, \\
        P(-)&= \bra{0}M_{-}\ket{0} = \abs{\braket{-}{0}}^{2}= \bra{1}M_{-}\ket{1} = \abs{\braket{-}{1}}^{2}=\frac{1}{2}.
    \end{aligned}
\end{equation}
The TP registers the measurement outcome as 0 if $\ket{+}$ is obtained and as 1 if $\ket{-}$ is obtained. Subsequently, all measurement outcomes are disclosed through the public channels. \\

\begin{table}[H]
\centering
\caption{\label{table 1} Different cases of the protocol depending on Alice's and Bob's chosen operations, along with their measurement outcomes, as well as the TP's measurement results.}
\resizebox{\textwidth}{!}{%
\begin{tabular}{ccccccc}
\hline\hline
 Case & Alice's operation & \thead{States obtained and \\ prepared by Alice} & Bob's operation & \thead{States obtained and \\ prepared by Bob} & \thead{State measured \\ by TP} & \thead{Probability of this \\ case happening} \\
\hline
 1 & MH & $\ket{0} \rightarrow \ket{+}$ & MH & $\ket{0} \rightarrow \ket{+}$ & $\ket{+}$ & $\cfrac{1}{16}$ \\
 2 & MH & $\ket{0} \rightarrow \ket{+}$ & MH & $\ket{1} \rightarrow \ket{-}$ & $\ket{-}$ & $\cfrac{1}{16}$ \\
 3 & MH & $\ket{0} \rightarrow \ket{+}$ & HM & $\ket{0} \rightarrow \ket{0}$ & $\ket{+}$ or $\ket{-}$ & $\cfrac{1}{8}$ \\
 4 & MH & $\ket{1} \rightarrow \ket{-}$ & MH & $\ket{0} \rightarrow \ket{+}$ & $\ket{+}$ & $\cfrac{1}{16}$ \\
 5 & MH & $\ket{1} \rightarrow \ket{-}$ & MH & $-\ket{1} \rightarrow -\ket{-}$ & $-\ket{-}$ & $\cfrac{1}{16}$ \\
 6 & MH & $\ket{1} \rightarrow \ket{-}$ & HM & $\ket{1} \rightarrow \ket{1}$ & $\ket{+}$ or $\ket{-}$ & $\cfrac{1}{8}$ \\
 7 & HM & $\ket{0} \rightarrow \ket{0}$ & MH & $\ket{0} \rightarrow \ket{+}$ & $\ket{+}$ & $\cfrac{1}{4}$ \\
 8 & HM & $\ket{0} \rightarrow \ket{0}$ & HM & $\ket{+} \rightarrow \ket{0}$ & $\ket{+}$ or $\ket{-}$ & $\cfrac{1}{8}$ \\
 9 & HM & $\ket{0} \rightarrow \ket{0}$ & HM & $\ket{+} \rightarrow \ket{1}$ & $\ket{+}$ or $\ket{-}$ & $\cfrac{1}{8}$ \\

\hline\hline
\end{tabular}%
}
\end{table}

\textbf{Step 5:} Once TP has revealed all the measurement results, Alice and Bob also reveal their chosen operations from steps 2 and 3 via the authenticated classical channel, without disclosing their measurement outcomes. Then, they discuss detecting eavesdroppers, ensuring TP's honesty, and obtaining a raw key. Depending on Alice and Bob's chosen operations and their measurement outcomes, nine cases emerge, as displayed in \hyperref[table 1]{Table 1}.

We see from \hyperref[table 1]{Table 1} that we have four situations as it follows:

\begin{itemize}
    \item \textbf{Situation 1:} In cases 1, 2, 4, and 5, when both Alice and Bob choose the \textbf{MH} operation in step 2 and step 3. We consistently note that TP's measurement result always matches the state sent by Bob. These cases serve as checks for the TP's honesty. Hence, to successfully pass the detection, the TP must consistently declare the same measurement result as Bob. Specifically, the TP must report a measurement result of 0 when Bob transmits the state $\ket{+}$ and 1 when Bob transmits the state $\ket{-}$.
    
    \item \textbf{Situation 2:} In cases 3, and 6, when Alice chooses the \textbf{MH} operation in step 2 and Bob selects the \textbf{HM} operation in step 3. We notice that Alice's measurement outcomes consistently match Bob's measurement outcomes. The qubits from these cases are utilized for generating the raw key. To check the validity of the key, Alice randomly selects a subset of these qubits and discloses their states. Bob subsequently reveals the states in the positions chosen by Alice. If there are no eavesdroppers and TP conducts the protocol honestly, the states of these disclosed qubits in corresponding positions should match. The unrevealed qubits will be utilized for generating the raw key.
    \item \textbf{Situation 3:}  In case 7, when Alice chooses the \textbf{HM} operation in step 2 and Bob selects the \textbf{MH} operation in step 3, both Alice and Bob, as well as the TP, obtain the measurement result of $0$. This case also serves as a check for the TP's honesty. Therefore, to successfully pass the detection, Alice and Bob must always obtain a measurement result of $0$, and the TP must consistently declare his measurement result to be the same.

    \item \textbf{Situation 4:} In cases 8, and 9, when both Alice and Bob select the \textbf{HM} operation in step 2, we notice here that TP should announce the measurement result as 0 or 1 with the same probability. However, Alice must consistently obtain a measurement result of 0; otherwise, the protocol is aborted.
\end{itemize}

\textbf{Step 6:} Alice and Bob calculate the error rate in each of the four situations described above. If the error rate exceeds a predetermined threshold in any of these situations, they will terminate the protocol. \\

\textbf{Step 7:} Alice and Bob will conduct error correction code (ECC) and privacy amplification (PA) on the raw key bits to obtain the final key.

\section{Security analysis}\label{sec3}

In this section, we analyze the security of the proposed MSQKD protocol. Our analysis focuses on common attacks typically examined in MSQKD protocols, including the measurement attack, faked states attack, and collective attack. We specifically consider the scenario where the TP acts as the attacker, as it possesses greater capabilities compared to other potential eavesdroppers. Therefore, any external eavesdropper's attack can be absorbed under the TP's attack. We assume that the probability of Alice and Bob selecting either the \textbf{MH} or \textbf{HM} operations in steps 2 and 3 is $\frac{1}{2}$.

\subsection{Measurement attack}\label{subsec3.1}

 The measurement attack entails the TP conducting measurements in bases other than the $X$ basis specified in step 4 of the protocol. This change aims at getting some side information about the raw key. In our analysis, we will consider that the TP may choose the $Z$ basis and the Breidbart basis as bases of measurement. Of course, the analysis is the same with any other basis.

\subsubsection{TP measures the received qubits in the $Z$ basis instead of the $X$ basis}\label{subsubsec3.1.1}
Initially, let's consider a scenario where the TP substitutes $X$ basis measurements with $Z$ basis measurements. However, in step 4, the TP must reveal his measurement outcomes to pass Alice and Bob's detection check. Since the TP is ignorant of Alice and Bob's chosen operations, reference to \hyperref[table 2]{Table 2} reveals that TP can only evade detection in cases 3, 6, 8, and 9, specifically when Bob selects the \textbf{HM} (Hadamard then Measure) operation in step 3. When Bob chooses the \textbf{MH} (Measure then Hadamard) operation, the TP needs to publish a measurement result similar to Bob's state. TP has a probability of $\frac{1}{2}$ of disclosing the correct measurement result, due to the fact of always receiving qubits in the $X$ basis:

\begin{equation}\label{eq13}
    \begin{aligned}
        P(0)&= \bra{+}M_{0}\ket{+} = \abs{\braket{0}{+}}^{2}= \bra{-}M_{0}\ket{-} = \abs{\braket{0}{-}}^{2}=\frac{1}{2}, \\
        P(1)&= \bra{+}M_{1}\ket{+} = \abs{\braket{1}{+}}^{2}= \bra{-}M_{1}\ket{-} = \abs{\braket{1}{-}}^{2}=\frac{1}{2}.
    \end{aligned}
\end{equation}
Consequently, if we refer to \hyperref[table 2]{Table 2}, TP has a total probability of $\frac{1}{2}\cdot(\frac{1}{16}+\frac{1}{16}+\frac{1}{16}+\frac{1}{16}+\frac{1}{4})=\frac{1}{4}$ of declaring the wrong result for a single qubit. Therefore, the probability of TP's attack being detected is computed as $1-(\frac{3}{4})^N$. With a sufficiently large $N$, the probability of TP's detection approaches certainty.

\begin{table}
\centering
\caption{\label{table 2} Different cases when TP performs a measurement in the $Z$-basis.}
\resizebox{\textwidth}{!}{%
\begin{tabular}{cccccccc}
\hline\hline
 Case & Alice's operation & \thead{State received \\ by Bob} & Bob's operation & \thead{State received \\ by TP} & \thead{State measured \\ by TP} & \thead{Probability of \\ being detected} & \thead{Probability of this \\ case happening}  \\
\hline
 1 & MH & $\ket{+}$ & MH & $\ket{+}$ & $\ket{0}$ or $\ket{1}$ & $\cfrac{1}{2}$ & $\cfrac{1}{16}$ \\
 2 & MH & $\ket{+}$ & MH & $\ket{-}$ & $\ket{0}$ or $-\ket{1}$ & $\cfrac{1}{2}$ & $\cfrac{1}{16}$ \\
 3 & MH & $\ket{+}$ & HM & $\ket{0}$ & $\ket{0}$ & $0$ & $\cfrac{1}{8}$ \\
 4 & MH & $\ket{-}$ & MH & $\ket{+}$ & $\ket{0}$ or $\ket{1}$ & $\cfrac{1}{2}$ & $\cfrac{1}{16}$ \\
 5 & MH & $\ket{-}$ & MH & $-\ket{-}$ & $-\ket{0}$ or $\ket{1}$ & $\cfrac{1}{2}$ & $\cfrac{1}{16}$ \\
 6 & MH & $\ket{-}$ & HM & $\ket{1}$ & $\ket{1}$ & $0$ & $\cfrac{1}{8}$ \\
 7 & HM & $\ket{0}$ & MH & $\ket{+}$ & $\ket{0}$ or $\ket{1}$ & $\cfrac{1}{2}$ & $\cfrac{1}{4}$ \\
 8 & HM & $\ket{0}$ & HM & $\ket{0}$ & $\ket{0}$ & $0$ & $\cfrac{1}{4}$ \\
 9 & HM & $\ket{0}$ & HM & $\ket{1}$ & $\ket{1}$ & $0$ & $\cfrac{1}{4}$ \\

\hline\hline
\end{tabular}%
}
\end{table}

\subsubsection{TP measures the received qubits in the Breidbart basis instead of the $X$ basis}\label{subsubsec3.1.2}

Next, we examine the scenario where the TP utilizes the Breidbart basis \cite{Dan2009} for measuring the received qubits. The reason behind this selection is that, as an intermediate basis between the $X$ and $Z$ bases, the Breidbart basis usually allows the attacker to obtain more side information while also decreasing the probability of being detected compared to the other bases mentioned in QKD protocols. The Breidbart basis is composed of the following two states:
\begin{equation}\label{eq14}
    \ket{\alpha_1} = \cos{\frac{\pi}{8}} \ket{0} + \sin{\frac{\pi}{8}} \ket{1}, \quad \ket{\alpha_2} = -\sin{\frac{\pi}{8}} \ket{0} + \cos{\frac{\pi}{8}} \ket{1}.
\end{equation}

Similar to the previous example, the TP must reveal his measurement result of 0 or 1 in step 4 to pass Alice's and Bob's detection test. We assume that the TP declares the measurement outcome as 0 when he obtains the state $\ket{\alpha_1}$; otherwise, he declares the measurement result as 1. The cases where the TP has a probability of publishing incorrect results are those where Bob selects the \textbf{MH} operation, and the TP needs to publish a measurement result similar to Bob's state. Regardless of whether Bob sent the state $\ket{+}$ or $\ket{-}$ in step 3, the probability that the TP announces an incorrect result is:
\begin{equation}\label{15}
    \abs{\braket{\alpha_1}{-}}^2 = \abs{\braket{\alpha_2}{+}}^2 = \frac{1}{2}(\cos{\frac{\pi}{8}}-\sin{\frac{\pi}{8}})^{2}.
\end{equation}
Hence, the probability of detecting the TP's attack is $1-(1-\frac{1}{4}(\cos{\frac{\pi}{8}}-\sin{\frac{\pi}{8}})^{2})^{N}$. As the value of $N$ increases, the probability of detecting the TP's attack approaches 1.

\subsection{Faked states attack}\label{subsec3.2}

The faked states attack involves the TP preparing qubits in states other than the state $\ket{+}$ as required in step 1 of the protocol. These states could be any single-qubit state or one part of an entangled state. Once again, the objective of this attack is to acquire side information about Alice's and Bob's secret key. In the following analysis, we will examine the security against this attack using two examples.

\subsubsection{TP prepares qubits in the $Z$ basis}

The first example entails the TP preparing and sending qubits in the $Z$ basis during step 1. In this scenario, we will investigate two situations: one where the TP substitutes the $X$ basis measurement for the received qubits in step 4 with a $Z$ basis measurement, and another where the TP retains the $X$ basis measurement. Note that the TP will declare the measurement result as 0 if they measure the state $\ket{0}$ in situation 1 and the state $\ket{+}$ in situation 2; otherwise, they will declare the measurement result as 1. 

But before delving into these, let's examine how a state in the $Z$ basis evolves after Alice's and Bob's operations. Considering the case where TP sends the state $\ket{0}$, upon receipt, Alice follows the protocol and chooses one of the two operations:

\begin{itemize}
    \item \textbf{MH \text{(Measure then Hadamard)}:} This involves measuring the received qubit in the $Z$-basis $\{\ket{0}, \ket{1}\}$,  generating a new qubit in the same state, applying a Hadamard operation on it, and then passing it to Bob.  When Alice chooses this operation, she measures the state $\ket{0}$ with certainty and transmits the state $\ket{+}=H\ket{0}$ to Bob.
    
    \item \textbf{HM \text{(Hadamard then Measure)}:} Here, the Hadamard operation is applied on the received qubit, then measured in the $Z$-basis. A new qubit is generated in the same state and sent to Bob. When applying the Hadamard operation on $\ket{0}$, Alice obtains the state $\ket{+}=H\ket{0}$. Thus, when she performs her measurement, she will obtain the state $\ket{0}$ or $\ket{1}$ with equal probability:
    \begin{equation}\label{eq16}
        \begin{aligned}
            P(0)&= \bra{+}M_{0}\ket{+} = \abs{\braket{0}{+}}^{2}=\frac{1}{2}, \\
            P(1)&= \bra{+}M_{1}\ket{+} = \abs{\braket{1}{+}}^{2}=\frac{1}{2}.
        \end{aligned}
    \end{equation}
    In the situation where Alice obtains the state $\ket{1}$, she will abort the protocol. This is because, when selecting the \textbf{HM} operation, she always expects to measure the state $\ket{0}$, as shown in \hyperref[Table 1]{table 1}. Therefore, TP's attack has a probability of $\frac{1}{4}$ of being detected in step 2.
\end{itemize}
From Alice's operation, she will either transmit the state $\ket{+}$ with a probability of $\frac{1}{2}$ or $\ket{0}$ with a probability of $\frac{1}{4}$.  After Bob receives a qubit from Alice, he performs the same operations as Alice did in step 2. If he performs the \textbf{MH} operation, he will either measure the state $\ket{0}$ or $\ket{1}$ with equal probability when he receives the state $\ket{+}$ from Alice. Alternatively, if he receives $\ket{0}$, he will measure $\ket{0}$ with certainty. if Bob chooses the \textbf{HM} operation, he will measure the state $\ket{0}$ with certainty if he receives $\ket{+}$ since $\ket{0}=H\ket{+}$; otherwise, he will obtain $\ket{0}$ or $\ket{1}$ with equal probability if he receives $\ket{0}$ since $\ket{+}=H\ket{0}$. This means that Bob will transmit one of the following states to TP: $\{\ket{+}, \ket{-}, \ket{0}, \ket{1}\}$.

\hyperref[table 3]{Table 3} displays all the possible cases. We can see that in cases 1, 2, 4, and 7, when Bob selects the \textbf{MH} operation, TP receives either $\ket{+}$ or $\ket{-}$. Consequently, if TP measures the received qubits in the $Z$ basis,  he has a probability of $\frac{1}{2}$ of announcing the wrong measurement outcome and thus being detected, as shown:
\begin{equation}\label{eq18}
    \begin{aligned}
        P(0)&= \bra{+}M_{0}\ket{+} = \abs{\braket{0}{+}}^{2}= \bra{-}M_{0}\ket{-} = \abs{\braket{0}{-}}^{2}=\frac{1}{2}, \\
        P(1)&= \bra{+}M_{1}\ket{+} = \abs{\braket{1}{+}}^{2}= \bra{-}M_{1}\ket{-} = \abs{\braket{1}{-}}^{2}=\frac{1}{2}.
    \end{aligned}
\end{equation}
This is because both Alice and Bob expect TP to declare the same measurement outcome as Bob when he selects the \textbf{MH} operation.

Additionally, we observe that the TP is detected half of the time, whether he chooses to measure the received qubits in the $Z$ basis or the $X$ basis when Alice selects the \textbf{HM} operation and obtains the measurement result $\ket{1}$ in cases 7, 8, and 9. This is because Alice always expects to obtain the measurement result $\ket{0}$ when she selects the \textbf{HM} operation.

By examining \hyperref[table 3]{Table 3}, the probability of TP being detected at each round while measuring in the $Z$ basis can be calculated as $\frac{1}{2}\cdot(\frac{1}{8}+\frac{1}{8}+\frac{1}{8})+\frac{1}{4}=\frac{7}{16}$. Thus, we can deduce that the TP's attack has a probability of $1-(\frac{9}{16})^{N}$ of being detected when the TP measures the received qubits in the $Z$ basis. 

When TP measures the received qubits in the $X$ basis, the probability of detection at each round is $\frac{1}{4}$. Therefore, the probability of detecting the attack when TP measures the received qubits in the $X$ basis is $1-(\frac{3}{4})^N$. These probabilities approach 1 when $N$ is sufficiently large. 

It is important to note that we analyze this scenario when the TP prepares and sends qubits in the state $\ket{0}$. However, we would obtain identical results if the TP prepared and sent qubits in the state $\ket{1}$.

\begin{table}[H]
\centering
\caption{\label{table 3} Different cases when TP sends qubits in the state $\ket{0}$ and performs a measurement in the $Z$-basis and $X$ basis respectively on the received qubits.}
\resizebox{\textwidth}{!}{%
\begin{tabular}{cccccccccc}
\hline\hline
 Case & \thead{Alice's \\ operation} & \thead{State received \\ by Bob} & \thead{Bob's \\ operation} & \thead{State received \\ by TP} & \thead{TP's Measurement result \\ in the $Z$ basis} & \thead{Probability of \\ being detected?} & \thead{TP's measurement result \\ in the $X$ basis} & \thead{Probability of \\ being detected?}& \thead{Probability of this \\ case happening} \\
\hline
 1 & MH & $\ket{+}$ & MH & $\ket{+}$ & $\ket{0}$ or $\ket{1}$ & $\cfrac{1}{2}$ & $\ket{+}$ & $0$ & $\cfrac{1}{8}$ \\
 2 & MH & $\ket{+}$ & MH & $\ket{-}$ & $\ket{0}$ or $-\ket{1}$ & $\cfrac{1}{2}$ & $\ket{-}$ & $0$ &  $\cfrac{1}{8}$ \\
 3 & MH & $\ket{+}$ & HM & $\ket{0}$ & $\ket{0}$ & $0$ & $\ket{+}$ or $\ket{-}$ & $0$ & $\cfrac{1}{4}$ \\
 4 & HM & $\ket{0}$ & MH & $\ket{+}$ & $\ket{0}$ or $\ket{1}$ & $\cfrac{1}{2}$ & $\ket{+}$ & $0$ & $\cfrac{1}{8}$ \\
 5 & HM & $\ket{0}$ & HM & $\ket{0}$ & $\ket{0}$ & $0$ & $\ket{+}$ or $\ket{-}$ & $0$ & $\cfrac{1}{16}$ \\
 6 & HM & $\ket{0}$ & HM & $\ket{1}$ & $\ket{1}$ & $0$ & $\ket{+}$ or $\ket{-}$ & $0$ & $\cfrac{1}{16}$ \\
 7 & HM & $\ket{1}$ & n/a & n/a & n/a & $1$ & n/a & $1$ & $\cfrac{1}{4}$ \\
\hline\hline
\end{tabular}%
}
\end{table}

\subsubsection{TP prepares qubits in the Bell basis}

In the second example, we consider the TP sending one particle of Bell states instead of the state $\ket{+}$ as required in the protocol. We remind that the Bell states can be represented as follows:

\begin{equation}\label{eq19}
    \begin{aligned}
        \ket{\phi^+} &= \frac{1}{\sqrt{2}} (\ket{00}+\ket{11})=\frac{1}{\sqrt{2}}(\ket{++}+\ket{--}), \\
        \ket{\phi^-} &= \frac{1}{\sqrt{2}} (\ket{00}-\ket{11})=\frac{1}{\sqrt{2}}(\ket{+-}+\ket{-+}), \\
        \ket{\psi^+} &= \frac{1}{\sqrt{2}} (\ket{01}+\ket{10})=\frac{1}{\sqrt{2}}(\ket{++}-\ket{--}), \\
        \ket{\psi^-} &= \frac{1}{\sqrt{2}} (\ket{01}-\ket{10})=\frac{1}{\sqrt{2}}(\ket{-+}-\ket{+-}), \\
    \end{aligned}
\end{equation}
Let's suppose the TP prepares the Bell state $\ket{\phi^+} = \frac{1}{\sqrt{2}}(\ket{00}+\ket{11})$ each time and sends one particle of this Bell state to Alice in step 1 while retaining the other particle. Upon receiving the qubit from TP, Alice follows the protocol and randomly performs one of two operations: \textbf{MH} (Measure then Hadamard) or \textbf{HM} (Hadamard then Measure) on her qubit. The state of Alice's qubit can be described by the reduced density operator $\rho_A$ as follows:

\begin{equation}\label{eq20}
    \begin{aligned}
        \rho_A &= \Tr_{TP}\left[\left(\cfrac{\ket{00}+\ket{11}}{\sqrt{2}}\right)\left(\cfrac{\bra{00}+\bra{11}}{\sqrt{2}}\right)\right] \\
               &= \cfrac{\ket{0}\bra{0}+\ket{1}\bra{1}}{2}.
    \end{aligned}
\end{equation}
This implies that when Alice selects the \textbf{MH} operation, the probability of obtaining either $\ket{0}$ or $\ket{1}$ upon measurement is $\frac{1}{2}$. Therefore, she will transmit the state $\ket{+}=H\ket{0}$ or $\ket{-}=H\ket{1}$ with equal probability to Bob. Note that if Alice transmits the state $\ket{+}$ the state of the quantum system is $\ket{+}\ket{0}$ and if Alice transmits the
the state $\ket{-}$ the quantum system is in the state $\ket{-}\ket{1}$.

Now, considering the scenario where Alice performs the \textbf{HM} operation, after applying the Hadamard operation, the state $\ket{\psi^+}$ becomes:

\begin{equation}\label{eq21}
    \begin{aligned}
        (H\otimes I)\ket{\psi^+} &= \frac{1}{\sqrt{2}}\left(\ket{+}\ket{0}+\ket{-}\ket{1}\right) \\
             &= \frac{1}              {\sqrt{2}}\left[\left(\cfrac{\ket{00}+\ket{10}}{\sqrt{2}}\right)+\left(\cfrac{\ket{01}-\ket{11}}{\sqrt{2}}\right)\right] \\
             &= \frac{1}{\sqrt{2}} \left[\ket{0}\left(\cfrac{\ket{0}+\ket{1}}{\sqrt{2}}\right) + \ket{1}\left(\cfrac{\ket{0}-\ket{1}}{\sqrt{2}}\right)\right] \\
             &= \frac{1}{\sqrt{2}}\left(\ket{0}\ket{+}+\ket{1}\ket{-}\right),
    \end{aligned}
\end{equation}
where $I$ is the identity operation. We can describe Alice's qubit by the reduced density operator $\rho_A$ as follow:

\begin{equation}\label{22}
    \begin{aligned}
        \rho_A &= \Tr_{TP}\left[\left(\cfrac{\ket{0}\ket{+}+\ket{1}\ket{-}}{\sqrt{2}}\right)\left(\cfrac{\bra{0}\bra{+}+\bra{1}\bra{-}}{\sqrt{2}}\right)\right] \\
               &= \cfrac{\ket{0}\bra{0}+\ket{1}\bra{1}}{2}.
    \end{aligned}
\end{equation}
Therefore, Alice has a probability of $\frac{1}{2}$ of measuring either the state $\ket{0}$ or $\ket{1}$. When Alice obtains $\ket{1}$ she will abort the protocol because when Alice selects the \textbf{HM} she always expects to obtain the state $\ket{0}$ upon measurement. When she transmits the state $\ket{0}$ to Bob, the state of the quantum system is $\ket{0}\ket{+}$.

Overall, Bob will receive the first qubit of one of the following states: $\{\ket{+}\ket{0}, \ket{-}\ket{1}, \ket{0}\ket{+}\}$. When Bob performs the \textbf{MH} operation and measures in the $Z$ basis, he obtains $\ket{0}$ or $\ket{1}$ with equal probability: 
\begin{equation}\label{eq23}
    \begin{aligned}
        P(0)&= \bra{+}M_{0}\ket{+} = \abs{\braket{0}{+}}^{2}= \bra{-}M_{0}\ket{-} = \abs{\braket{0}{-}}^{2}=\frac{1}{2}, \\
        P(1)&= \bra{+}M_{1}\ket{+} = \abs{\braket{1}{+}}^{2}= \bra{-}M_{1}\ket{-} = \abs{\braket{1}{-}}^{2}=\frac{1}{2},
    \end{aligned}
\end{equation}
regardless of whether he receives $\ket{+}$ or $\ket{-}$, and obtain $\ket{0}$ with certainty if he receives $\ket{0}$. Thus, after performing Hadamard, Bob will either transmit $\ket{+}=H\ket{0}$ or $\ket{-}=H\ket{1}$ based on his measurement outcome. This means that the state $\ket{+}\ket{0}$ evolves into either $\ket{+}\ket{0}$ or $\ket{-}\ket{0}$ with equal probability. The state $\ket{-}\ket{1}$ evolves into either $\ket{+}\ket{1}$ or $-\ket{-}\ket{1}$ with equal probability. Finally, the state $\ket{0}\ket{+}$ evolves into $\ket{+}\ket{+}$ with certainty. 

Now, considering when Bob chooses the \textbf{HM} operation, after the initial Hadamard, depending on Alice's sent state, Bob ends up with one of the following states:
\begin{equation}\label{eq24}
    \ket{0}=H\ket{+}, \quad \ket{1}=H\ket{-}, \quad \ket{+}=H\ket{0}.
\end{equation}
After measuring in the $Z$ basis, he will transmit $\ket{0}$ or $\ket{1}$ based on his outcome. Thus, $\ket{+}\ket{0}$ evolves into $\ket{0}\ket{0}$, $\ket{-}\ket{1}$ evolves into $\ket{1}\ket{1}$, and $\ket{0}\ket{+}$ will evolve into either $\ket{0}\ket{+}$ or $\ket{-}\ket{0}$ with equal probability.

Upon receipt, the TP conducts a two-qubit measurement on the retained particle and the received qubit. Similar to the previous scenario, we analyze two situations: one where the TP performs a Bell measurement, and another where the TP performs a measurement in the basis $\{\ket{00}, \ket{01}, \ket{10}, \ket{11}\}$. Of course, TP can perform any two-qubit measurement on the qubit sent by Bob and the preserved qubit. We just select those two because they are the most common.
All cases based on Alice and Bob's choice of operations and their measurement results in both situations are respectively displayed in \hyperref[table 4]{table 4} and \hyperref[table 5]{table 5}.

As shown in \hyperref[table 4]{table 4}, when the TP obtains the state $\ket{\phi^+}$, he cannot acquire any information. However, if the TP obtains $\ket{\phi^-}$, the only inference possible is that he is not in cases 7 and 10, when Alice and Bob respectively choose the \textbf{HM} operation and the \textbf{MH} operation.  Therefore, it is certain that the TP cannot extract any information about the raw key when performing a Bell measurement. Assuming the TP announces the measurement result as 0 when he obtains the states $\{\ket{\phi^+}, \ket{\psi^+}\}$; otherwise, the TP announces the result as 1. \hyperref[table 4]{table 4} reveals that the TP has a probability of $\frac{1}{2}$ of declaring the wrong measurement outcome in cases 1, 2, 4, and 5 when both Alice and Bob select the \textbf{MH} operation. Furthermore, the TP is detected with certainty in case 10 when Alice selects the \textbf{HM} operation and obtains the measurement $\ket{1}$ for her qubit. Consequently, the probability of detection at each round can be computed as $\cfrac{1}{2}\cdot(\frac{1}{16}+\frac{1}{16}+\frac{1}{16}+\frac{1}{16})+ \frac{1}{4}=\frac{3}{8}$. Thus, the probability of detecting the TP's attack when choosing to perform a Bell measurement can be deduced as $1-(\frac{5}{8})^N$. As $N$ increases, this probability approaches 1.

In the scenario where the TP chooses to perform the measurement in the basis $\{\ket{00}, \ket{01}, \ket{10}, \ket{11}\}$, we assume the TP declares the measurement result as 0 when he obtains the states ${\ket{00}, \ket{01}}$; otherwise, they announce the measurement result as 1. From \hyperref[table 5]{table 5}, it is evident that although the TP manages to acquire information about the raw key in cases 3 and 6, he still has a probability of $\frac{1}{2}$ of announcing the wrong result in cases 1, 2, 4, and 5. Moreover, TP is detected with certainty in case 10 when Alice selects the \textbf{HM} operation and obtains the measurement $\ket{1}$ for her qubit. Thus, the probability of detection at each round can be calculated as $\frac{1}{2}\cdot(6\cdot\frac{1}{16})=\frac{7}{16}$. Therefore, TP's attack when performing measurements in the basis $\{\ket{00}, \ket{01}, \ket{10}, \ket{11}\}$ can be easily computed as $1-(\frac{9}{16})^N$, with the probability approaching 1 as $N$ increases.

\begin{table}[H]
\centering
\caption{\label{table 4} Different cases when TP uses one particle of the Bell state as the faked state and conducts a measurement in the Bell Basis.}
\resizebox{\textwidth}{!}{%
\begin{tabular}{cccccccc}
\hline\hline
 Case & \thead{Alice's \\ operation} & \thead{State received \\ by Bob} & \thead{Bob's \\ operation} & \thead{State received \\ by TP} & TP's Measurement result & \thead{Probability of \\ being detected} &  \thead{Probability of this \\ case happening} \\
\hline
 1 & MH & $\ket{+}\ket{0}$ & MH & $\ket{+}\ket{0}$ & $\ket{\phi^+}$ or $\ket{\phi^-}$ or $\ket{\psi^+}$ or $\ket{\psi^-}$ & $\cfrac{1}{2}$ & $\cfrac{1}{16}$ \\
 2 & MH & $\ket{+}\ket{0}$ & MH & $\ket{-}\ket{0}$ & $\ket{\phi^+}$ or $\ket{\phi^-}$ or $\ket{\psi^+}$ or $\ket{\psi^-}$ & $\cfrac{1}{2}$ & $\cfrac{1}{16}$ \\
 3 & MH & $\ket{+}\ket{0}$ & HM & $\ket{00}$ & $\ket{\phi^+}$ or $\ket{\phi^-}$ & $0$ & $\cfrac{1}{8}$ \\
 4 & MH & $\ket{-}\ket{1}$ & MH & $\ket{+}\ket{1}$ & $\ket{\phi^+}$ or $\ket{\phi^-}$ or $\ket{\psi^+}$ or $\ket{\psi^-}$ & $\cfrac{1}{2}$ & $\cfrac{1}{16}$ \\
 5 & MH & $\ket{-}\ket{1}$ & MH & $-\ket{-}\ket{1}$ & $\ket{\phi^+}$ or $\ket{\phi^-}$ or $\ket{\psi^+}$ or $\ket{\psi^-}$ & $\cfrac{1}{2}$ & $\cfrac{1}{16}$ \\
 6 & MH & $\ket{-}\ket{1}$ & HM & $\ket{11}$ & $\ket{\phi^+}$ or $\ket{\phi^-}$ & $0$ & $\cfrac{1}{8}$ \\
 7 & HM & $\ket{0}\ket{+}$ & MH & $\ket{++}$ & $\ket{\phi^+}$ or $\ket{\psi^+}$ & $0$ & $\cfrac{1}{8}$ \\
 8 & HM & $\ket{0}\ket{+}$ & HM & $\ket{0}\ket{+}$ & $\ket{\phi^+}$ or $\ket{\phi^-}$ or $\ket{\psi^+}$ or $\ket{\psi^-}$ & $0$ & $\cfrac{1}{16}$ \\
 9 & HM & $\ket{0}\ket{+}$ & HM & $\ket{1}\ket{+}$ & $\ket{\phi^+}$ or $\ket{\phi^-}$ or $\ket{\psi^+}$ or $\ket{\psi^-}$ & $0$ & $\cfrac{1}{16}$ \\
 10 & HM & $\ket{1}\ket{-}$ & n/a & n/a & n/a  & $1$ & $\cfrac{1}{4}$ \\
\hline\hline
\end{tabular}%
}
\end{table}

\begin{table}[H]
\centering
\caption{\label{table 5} Different cases when TP uses one particle of the Bell state as the faked state and conducts a measurement in the Basis $\{\ket{00}, \ket{01}, \ket{10}, \ket{11} \}$.}
\resizebox{\textwidth}{!}{%
\begin{tabular}{cccccccc}
\hline\hline
 Case & \thead{Alice's \\ operation} & \thead{State received \\ by Bob} & \thead{Bob's \\ operation} & \thead{State received \\ by TP} & TP's Measurement result & \thead{Probability of \\ being detected} &  \thead{Probability of this \\ case happening} \\
\hline
 1 & MH & $\ket{+}\ket{0}$ & MH & $\ket{+}\ket{0}$ & $\ket{00}$ or $\ket{10}$ & $\cfrac{1}{2}$ & $\cfrac{1}{16}$ \\
 2 & MH & $\ket{+}\ket{0}$ & MH & $\ket{-}\ket{0}$ & $\ket{00}$ or $\ket{10}$ & $\cfrac{1}{2}$ & $\cfrac{1}{16}$ \\
 3 & MH & $\ket{+}\ket{0}$ & HM & $\ket{00}$ & $\ket{00}$ & $0$ & $\cfrac{1}{8}$ \\
 4 & MH & $\ket{-}\ket{1}$ & MH & $\ket{+}\ket{1}$ & $\ket{01}$ or $\ket{11}$ & $\cfrac{1}{2}$ & $\cfrac{1}{16}$ \\
 5 & MH & $\ket{-}\ket{1}$ & MH & $-\ket{-}\ket{1}$ & $\ket{01}$ or $\ket{11}$ & $\cfrac{1}{2}$ & $\cfrac{1}{16}$ \\
 6 & MH & $\ket{-}\ket{1}$ & HM & $\ket{11}$ & $\ket{11}$ & $0$ & $\cfrac{1}{8}$ \\
 7 & HM & $\ket{0}\ket{+}$ & MH & $\ket{++}$ & $\ket{00}$ or $\ket{01}$ or $\ket{10}$ or $\ket{11}$ & $\cfrac{1}{2}$ & $\cfrac{1}{8}$ \\
 8 & HM & $\ket{0}\ket{+}$ & HM & $\ket{0}\ket{+}$ & $\ket{00}$ or $\ket{01}$ & $0$ & $\cfrac{1}{16}$ \\
 9 & HM & $\ket{0}\ket{+}$ & HM & $\ket{1}\ket{+}$ & $\ket{10} \text{or} \ket{11}$ & $0$ & $\cfrac{1}{16}$ \\
 10 & HM & $\ket{1}\ket{-}$ & n/a & n/a & n/a  & $1$ & $\cfrac{1}{4}$ \\
 
\hline\hline
\end{tabular}%
}
\end{table}

\subsection{Collective attack}\label{subsec3.3}

We define the collective attack as follows \cite{Iqbal2020, Krawec2015, Tsai2021}:  

\begin{itemize}
    \item TP's strategy consists of entangling an ancillary qubit with each quantum system sent between participants in the quantum channels. TP then proceeds by measuring those ancillary qubits to learn useful information about Alice's and Bob's secret key.
    \item TP uses the same strategy at each iteration of the protocol.
    \item TP is free to keep his ancillary qubits in a quantum memory and measure them at an ulterior time.
\end{itemize}

The protocol employs three distinct one-way quantum channels, allowing TP to conduct three separate $U_i$ operations, such that $U^{\dagger}_{i}U_{i}= I$, where $I$ is the identity matrix, on the transmitted qubits and their associated ancillary qubits. Note that the $U_i$ operations conducted on the same quantum channel remain identical. Additionally, TP retains the ancillary qubits even after Alice and Bob complete all classical post-processing operations.\\
The goal of this study is to demonstrate that TP cannot gain any valuable information about Alice's and Bob's secret key without disturbing the original quantum systems and therefore has a nonzero probability of being detected during the honesty checks.
There are two strategies that the TP can employ to attack the proposed protocol. In the first strategy, TP utilizes fresh ancillary qubits for each $U_i$, and the states of these ancillary qubits remain distinguishable after $U_i$ is performed. The second strategy involves performing different $U_i$ operations on the same ancillary qubit instead of using fresh ones for each channel. As mentioned by Chen et al. \cite{Chen2021}, the first strategy may be considered even more realistic, given the spatial separation of the channels, making it considerably challenging for an adversary to possess a single quantum ancilla in practice.

\subsubsection{Attacking strategy 1}\label{subsubsec3.3.1}

In this attack strategy, TP employs new ancillary qubits for each $U_i$, ensuring that the states of these ancillary qubits remain distinguishable after each operation $U_i$. TP retains these ancillary qubits even after Alice and Bob finish all classical post-processing operations.

As a first step, TP prepares the state $\ket{+}$ in step 1, followed by executing the $U_1$ operation on the state $\ket{+}$ and his ancillary qubit initially in the state $\ket{e_1}$, resulting in:

\begin{equation}
\label{eq25}
    U_{1}(\ket{+}\otimes\ket{e_1}) = a_{0}\ket{0}\ket{f_0} + a_{1}\ket{1}\ket{f_1},
\end{equation}
such that $\abs{a_0}^{2}+\abs{a_1}^{2}=1$, and $\{\ket{f_0}, \ket{f_1}\}$ is the orthogonal basis that TP can distinguish. Then TP sends the first qubit to Alice. Upon receiving the qubit, Alice performs one of two operations: \textbf{(MH)} she measures the received qubit in the $Z$ basis, prepares a new qubit corresponding to the measurement result, then performs a Hadamard operation before sending it to Bob; and \textbf{(HM)} she performs a Hadamard operation on the received qubit, measures it in the $Z$ basis, prepares a new qubit corresponding to the measurement result, and sends it to Bob.

In \hyperref[table 6]{Table 6}, we observe the different measurement outcomes of Alice depending on her chosen operation as well as the state of the quantum system that consists of Alice's qubit and TP's ancilla. When Alice chooses the \textbf{HM} operation, she anticipates consistently obtaining the state $\ket{0}$ upon measurement according to \hyperref[table 1]{Table 1}. However, when the quantum system is in the state $a_{0}\ket{0}\ket{f_0}+a_{1}\ket{1}\ket{f_1}$, we see from \hyperref[table 6]{Table 6} that Alice has a probability of $\frac{1}{2}$ of obtaining the state $\ket{1}$ when she selects the \textbf{HM} operation, leading for the protocol to abort.

\begin{table}[H]
\caption{\label{table 6} The state of the quantum system (QS) after Alice's operations.}
\tiny
\resizebox{\textwidth}{!}{%
\begin{tabular}{ccccc}
\hline\hline
 Case & Alice's operation & \thead{\tiny State prepared \\ \tiny  by Alice} & \thead{\tiny State of the QS after \\ \tiny Alice's operation} & \thead{\tiny Probability of this \\ \tiny case happening} \\
\hline
 1 & MH & $\ket{+}$ & $\ket{+}\ket{f_0}$ & $\cfrac{1}{2} \abs{a_0}^2$ \\
 2 & MH & $\ket{-}$ & $\ket{-}\ket{f_1}$ & $\cfrac{1}{2} \abs{a_1}^2$ \\
 3 & HM & $\ket{0}$ & $\ket{0}(a_{0}\ket{f_0}+a_{1}\ket{f_1})$  & $\cfrac{1}{4}$ \\
 4 & HM & $\ket{1}$ & $\ket{1}(a_{0}\ket{f_0}-a_{1}\ket{f_1})$  & $\cfrac{1}{4}$ \\

\hline\hline
\end{tabular}%
}

\end{table}

To avoid this situation, TP has to set the incorrect item, namely $a_{0}\ket{f_0}-a_{1}\ket{f_1}$, to be a zero vector. This would entail $a_{0}\ket{f_0}=a_{1}\ket{f_1}$, which contradicts the orthogonality of $\ket{f_0}$ and $\ket{f_1}$, unless TP sets $a_0 = a_1 = 0$. However, this contradicts the equation $\abs{a_0}^{2}+\abs{a_1}^{2}=1$. In the scenario where TP cannot distinguish $a_{0}\ket{f_0}$ from $a_{1}\ket{f_1}$, he is unable to measure his ancillary qubit to extract information about Alice's measurement result. Consequently, there exists no $U_1$ operation that enables TP to acquire useful key information without having a non-zero probability of being detected.

We notice from \hyperref[table 6]{Table 6} that the quantum system consisting of Alice's qubit and TP's ancillary qubit is separable. Additionally, when TP executes his attack on the second channel, he will apply the $U_2$ operation to the qubit sent by Alice and his second ancillary qubit.  Therefore, it is equivalent to continue the analysis of TP's attack on the second channel by considering either the entire quantum system or focusing solely on the quantum system composed of the intercepted qubit and TP's second ancillary qubit, without considering the state of the first ancillary. For simplicity, we choose the latter option.

Following Alice's operations and the transmission of the state to Bob, TP intercepts the qubit and applies the $U_2$ operation to it and his ancillary qubit, initially in state $\ket{e_2}$. The $U_2$ operation is defined as follows:

\begin{equation}\label{eq26}
\begin{aligned}
  U_{2}(\ket{+}\otimes\ket{e_2}) &= b_{0}\ket{0}\ket{g_0} + b_{1}\ket{1}\ket{g_1}, \\
  U_{2}(\ket{-}\otimes\ket{e_2}) &= c_{0}\ket{0}\ket{h_0} + c_{1}\ket{1}\ket{h_1},
\end{aligned}
\end{equation}
such that $\abs{b_0}^{2}+\abs{b_1}^{2}=1$, $\abs{c_0}^{2}+\abs{c_1}^{2}=1$ and $\{\ket{g_0}, \ket{g_1}\}$, $\{\ket{h_0}, \ket{h_1}\}$ are orthogonal bases which TP can distinguish. By linearity, when  TP conducts his attack on Alice's qubit in the case she selected the \textbf{HM} operation, $U_2$ is represented as:

\begin{equation}\label{eq27}
    U_{2}(\ket{0}\otimes\ket{e_2}) = \frac{1}{\sqrt{2}}\{\ket{0}(b_{0}\ket{g_0}+c_{0}\ket{h_0}) + \ket{1}(b_{1}\ket{g_1}+c_{1}\ket{h_1})\}.
\end{equation}
TP then proceeds by sending the first qubit to Bob. Upon receiving it, Bob performs operations similar to those of Alice, but TP becomes the recipient of the state. Referring to \hyperref[table 7]{Table 7}, when Alice chooses the \textbf{MH} operation, she transmits either $\ket{+}$ or $\ket{-}$ with equal probability, leading to four possible outcomes after Bob's operation for each state. Conversely, when Alice selects the \textbf{HM} operation, she transmits the state $\ket{0}$, also resulting in four possible outcomes after Bob's operation. These outcomes are illustrated in \hyperref[table 7]{Table 7}.

\begin{table}
\caption{\label{table 7} The state of the quantum system (QS) after Bob's operations.}
\resizebox{\textwidth}{!}{%
\begin{tabular}{ccccccc}
\hline\hline
 Case & Alice's operation & \thead{State prepared \\ by Alice} & Bob's operation & \thead{State prepared \\ by Bob} & State of the QS & \thead{Probability of this \\ case happening} \\
\hline
 1 & MH & $\ket{+}$ & MH & $\ket{+}$ & State \Romannum{1} & $\cfrac{1}{8} \abs{b_{0}}^2$ \\
 2 & MH & $\ket{+}$ & MH & $\ket{-}$ & State \Romannum{2} & $\cfrac{1}{8} \abs{b_{1}}^2$ \\
 3 & MH & $\ket{+}$ & HM & $\ket{0}$ & State \Romannum{3} & $\cfrac{1}{16}$ \\
 4 & MH & $\ket{+}$ & HM & $\ket{1}$ & State \Romannum{4} & $\cfrac{1}{16}$ \\
 
 5 & MH & $\ket{-}$ & MH & $\ket{+}$ & State \Romannum{5} & $\cfrac{1}{8} \abs{c_{0}}^2$ \\
 6 & MH & $\ket{-}$ & MH & $\ket{-}$ & State \Romannum{6} & $\cfrac{1}{8} \abs{c_{1}}^2$ \\
 7 & MH & $\ket{-}$ & HM & $\ket{0}$ & State \Romannum{7} & $\cfrac{1}{16}$ \\
 8 & MH & $\ket{-}$ & HM & $\ket{1}$ & State \Romannum{8} & $\cfrac{1}{16}$ \\
 
 9 & HM & $\ket{0}$ & MH & $\ket{+}$ & State \Romannum{9} & $\cfrac{1}{8}(\abs{b_0}^{2}+\abs{c_0}^{2})$ \\
 10 & HM & $\ket{0}$ & MH & $\ket{-}$ & State \Romannum{10} & $\cfrac{1}{8}(\abs{b_1}^{2}+\abs{c_1}^{2})$ \\
 11 & HM & $\ket{0}$ & HM & $\ket{0}$ & State \Romannum{11} & $\cfrac{1}{8}$ \\
 12 & HM & $\ket{0}$ & HM & $\ket{1}$ & State \Romannum{12} & $\cfrac{1}{8}$ \\

\hline\hline
\end{tabular}%
}
\footnotesize
State \Romannum{1}: $\ket{+}\ket{g_0}$ \\
State \Romannum{2}: $\ket{-}\ket{g_1}$ \\
State \Romannum{3}: $\ket{0}(b_{0}\ket{g_0}+b_{1}\ket{g_1})$ \\
State \Romannum{4}: $\ket{1}(b_{0}\ket{g_0}-b_{1}\ket{g_1})$ \\
State \Romannum{5}: $\ket{+}\ket{h_0}$ \\
State \Romannum{6}: $\ket{-}\ket{h_1}$ \\
State \Romannum{7}: $\ket{0}(c_{0}\ket{h_0}+c_{1}\ket{h_1})$ \\
State \Romannum{8}: $\ket{1}(c_{0}\ket{h_0}-c_{1}\ket{h_1})$ \\
State \Romannum{9}: $\cfrac{1}{\sqrt{p_9}}\{\ket{+}(b_{0}\ket{g_0}+c_{0}\ket{h_0})\},$ such as $p_{9}=\cfrac{1}{2}(\abs{b_0}^{2}+\abs{c_0}^{2})$ \\
State \Romannum{10}: $\cfrac{1}{\sqrt{p_{10}}}\{\ket{-}(b_{1}\ket{g_1}+c_{1}\ket{h_1})\},$ such as $p_{10}=\cfrac{1}{2}(\abs{b_1}^{2}+\abs{c_1}^{2})$  \\
State \Romannum{11}: $\ket{0}(b_{0}\ket{g_0}+c_{0}\ket{h_0}+b_{1}\ket{g_1}+c_{1}\ket{h_1})$ \\
State \Romannum{12}: $\ket{1}(b_{0}\ket{g_0}+c_{0}\ket{h_0}-b_{1}\ket{g_1}-c_{1}\ket{h_1})$. \\

\end{table}

First, let's focus on the cases where Alice chooses the \textbf{HM} operation and Bob selects the \textbf{MH} operation. In this scenario, if we refer to \hyperref[table 1]{Table 1}, both Alice and Bob should consistently measure the state $\ket{0}$, otherwise, the protocol is aborted. The question now is whether TP can ensure that Bob measures $\ket{0}$. To achieve this, TP must avoid case 10 in \hyperref[table 7]{Table 7} and therefore set the incorrect term $b_{1}\ket{g_1}+c_{1}\ket{h_1}$ to a zero vector. This requires meeting the condition:

\begin{equation}
\label{eq27}
         b_{1}\ket{g_1}+c_{1}\ket{h_1} = 0.
\end{equation}

Next, we take a look at the situation when Alice opts for the \textbf{MH} operation and Bob selects the \textbf{HM} operation. If we refer again to \hyperref[table 1]{Table 1}, we know that the qubits involved in this scenario serve to establish the raw key due to the consistent correlation in Alice's and Bob's measurement outcomes. However, from \hyperref[table 7]{Table 7}, it's evident that TP's interference disrupts this pattern. Indeed, Alice and Bob observe anti-correlated results half of the time. For instance, in case 4, when Alice prepares the state $\ket{+}$ (i.e., obtaining $\ket{0}$ during her measurement), Bob records $\ket{1}$. Similarly, in case 7, when Alice prepares $\ket{-}$ (i.e., obtaining $\ket{1}$ during her measurement), Bob records $\ket{0}$. Alice and Bob identify these anti-correlations when they reveal the states of a random subset of these qubits to check the validity of the key. These discrepancies are indicative of the interference introduced by TP in the protocol's execution, prompting them to abort the protocol.

In order for TP to successfully pass the public discussion, he has to set the terms $b_{0}\ket{g_0}-b_{1}\ket{g_1}$ and $c_{0}\ket{h_0}+c_{1}\ket{h_1}$ as a zero vector. To avoid case 4, the following conditions must be met:

\begin{equation}
\label{eq28}
    b_{0}\ket{g_0} - b_{1}\ket{g_1} = 0.
\end{equation}
Similarly, to evade case 7, the conditions are:
\begin{equation}
\label{eq29}
    c_{0}\ket{h_0} + c_{1}\ket{h_1} = 0.
\end{equation}
So, \hyperref[eq27]{equation (27)} outlines the condition TP must meet for Bob to consistently get the correct measurement of $\ket{0}$ when Alice and Bob choose the operations \textbf{HM} and \textbf{MH}, respectively. \hyperref[eq28]{Equation (28)} and \hyperref[eq29]{equation (29)} specify the conditions to avoid the anti-correlation between Alice's and Bob's results when establishing the raw key.

Let's look at these conditions together. If we set $b_{1}\ket{g_1}-c_{1}\ket{h_1} = 0$ it implies that $b_{1}\ket{g_1}= -c_{1}\ket{h_1}$, making it impossible for TP to distinguish $b_{1}\ket{g_1}$ and $-c_{1}\ket{h_1}$. Combining these conditions with those in \hyperref[eq28]{equation (28)} and \hyperref[eq29]{equation (29)}, namely ($b_{0}\ket{g_0} - b_{1}\ket{g_1} = 0$ and $c_{0}\ket{h_0} + c_{1}\ket{h_1} = 0$), results in 
\begin{equation}\label{eq30}
    b_{0}\ket{g_0}=b_{1}\ket{g_1}=c_{0}\ket{h_0}=-c_{1}\ket{h_1}.
\end{equation}
This situation leaves TP unable to distinguish between $b_{0}\ket{g_0}$, $b_{1}\ket{g_1}$, $c_{0}\ket{h_0}$, and $-c_{1}\ket{h_1}$, which violates the orthogonality and distinguishability of the bases $\{\ket{g_0}, \ket{g_1}\}$ and $\{\ket{h_0}, \ket{h_1}\}$, unless TP sets all coefficients to zero: $b_0 = b_1 = c_0 = c_1 =0$. However, this contradicts the equations $\abs{b_0}^{2}+\abs{b_1}^{2}=1$ and $\abs{c_0}^{2}+\abs{c_1}^{2}=1$. Consequently, TP cannot measure his ancillary qubit to gain insight into Bob's measurement results, rendering him unable to obtain any useful information about Alice's and Bob's secret key. Thus, there is no $U_2$ operation that would allow TP to obtain any information about Alice's and Bob's secret key without being detected.

Looking at \hyperref[table 7]{Table 7}, similar to the attack on the first channel, we observe that Bob's qubit and TP's second ancillary qubit are not entangled. Moreover, when TP executes his attack on the third channel, he will apply the $U_3$ operation to the qubit sent by Bob and his third ancillary qubit.  As a result, we can analyze TP's attack on the second channel without considering the state of the second ancillary qubit. Instead, we focus solely on the quantum system composed of the intercepted qubit and TP's third ancillary qubit. After Bob's operations, he sends the qubit to TP. Upon receipt, TP performs the $U_3$ operation on this qubit and his ancillary qubit, initially in the state $\ket{e_3}$. The $U_3$ operation is defined as follows:
\begin{equation}\label{eq31}
    \begin{aligned}
         U_{3}(\ket{+}\otimes\ket{e_3}) = d_{0}\ket{0}\ket{i_0} + d_{1}\ket{1}\ket{i_1}, \\
         U_{3}(\ket{-}\otimes\ket{e_3}) = e_{0}\ket{0}\ket{j_0} + e_{1}\ket{1}\ket{j_1},
    \end{aligned}
\end{equation}
such that $\abs{d_0}^{2}+\abs{d_1}^{2}=1$, $\abs{e_0}^{2}+\abs{e_1}^{2}=1$ and $\{\ket{i_0}, \ket{i_1}\}$, $\{\ket{j_0}, \ket{j_1}\}$ are orthogonal bases which TP can distinguish. By linearity, this yields: 
\begin{equation}
\label{eq32}
    \begin{aligned}
         U_{3}(\ket{0}\otimes\ket{e_3}) = \frac{1}{\sqrt{2}}\{\ket{0}(d_{0}\ket{i_0}+e_{0}\ket{j_0}) + \ket{1}(d_{1}\ket{i_1}+e_{1}\ket{j_1})\}, \\
         U_{3}(\ket{1}\otimes\ket{e_3}) = \frac{1}{\sqrt{2}}\{\ket{0}(d_{0}\ket{i_0}-e_{0}\ket{j_0}) + \ket{1}(d_{1}\ket{i_1}-e_{1}\ket{j_1})\}.
    \end{aligned}
\end{equation}
Upon completion of the $U_3$ operation, the state of the quantum system changes, as illustrated in \hyperref[table 8]{Table 8}.

\begin{table}
\caption{\label{table 8} The state of the quantum system (QS) after TP performs the $U_3$ operation.}
\resizebox{\textwidth}{!}{%
\begin{tabular}{ccccccc}
\hline\hline
 Case & Alice's operation & \thead{State prepared \\ by Alice} & Bob's operation & \thead{State prepared \\ by Bob} & State of the QS & \thead{Probability of this \\ case happening} \\
\hline
 1 & MH & $\ket{+}$ & MH & $\ket{+}$ & State \Romannum{1} & $\cfrac{1}{16}$ \\
 2 & MH & $\ket{+}$ & MH & $\ket{-}$ & State \Romannum{2} & $\cfrac{1}{16}$ \\
 3 & MH & $\ket{+}$ & HM & $\ket{0}$ & State \Romannum{3} & $\cfrac{1}{8}$ \\
 
 4 & MH & $\ket{-}$ & MH & $\ket{+}$ & State \Romannum{1} & $\cfrac{1}{16}$ \\
 5 & MH & $\ket{-}$ & MH & $\ket{-}$ & State \Romannum{2} & $\cfrac{1}{16}$ \\
 6 & MH & $\ket{-}$ & HM & $\ket{1}$ & State \Romannum{4} & $\cfrac{1}{8}$ \\
 
 7 & HM & $\ket{0}$ & MH & $\ket{+}$ & State \Romannum{1} & $\cfrac{1}{4}$ \\
 8 & HM & $\ket{0}$ & HM & $\ket{0}$ & State \Romannum{3} & $\cfrac{1}{8}$ \\
 9 & HM & $\ket{0}$ & HM & $\ket{1}$ & State \Romannum{4} & $\cfrac{1}{8}$ \\

\hline\hline
\end{tabular}%
}
\footnotesize
State \Romannum{1}: $d_{0}\ket{0}\ket{i_0}+d_{1}\ket{1}\ket{i_1}$ \\
State \Romannum{2}: $e_{0}\ket{0}\ket{j_0}+e_{1}\ket{1}\ket{j_1}$ \\
State \Romannum{3}: $\cfrac{1}{\sqrt{2}}\{\ket{0}(d_{0}\ket{i_0}+e_{0}\ket{j_0}) + \ket{1}(d_{1}\ket{i_1}+e_{1}\ket{j_1})\}$ \\
State \Romannum{4}: $\cfrac{1}{\sqrt{2}}\{\ket{0}(d_{0}\ket{i_0}-e_{0}\ket{j_0}) + \ket{1}(d_{1}\ket{i_1}-e_{1}\ket{j_1})\}$

\end{table}

Subsequently, TP measures the received qubit in the $X$ basis according to the protocol. If he obtains the measurement result $\ket{+}$, he publishes 0; otherwise, he publishes 1. To pass the detection, TP must obtain the same measurement result as the state sent by Bob when Bob chooses the \textbf{MH} operation. We rewrite state \Romannum{1} and state \Romannum{2} in \hyperref[table 8]{Table 8} as follows:
\begin{equation}\label{eq33}
    \begin{aligned}
         \ket{\psi_{\Romannum{1}}} = \frac{1}{\sqrt{2}}\{\ket{+}(d_{0}\ket{i_0}+d_{1}\ket{i_1}) + \ket{-}(d_{0}\ket{i_0}-d_{1}\ket{i_1})\}, \\
         \ket{\psi_{\Romannum{2}}} = \frac{1}{\sqrt{2}}\{\ket{+}(e_{0}\ket{j_0}+e_{1}\ket{j_1}) + \ket{-}(e_{0}\ket{j_0}-e_{1}\ket{j_1})\}.
    \end{aligned}
\end{equation}
For TP to pass the detection, he must set the incorrect items as zero vectors. That is, TP will set:

\begin{equation}\label{eq34}
    \begin{aligned}
        d_{0}\ket{i_0}-d_{1}\ket{i_1} = 0, \\
        e_{0}\ket{j_0}+e_{1}\ket{j_1} = 0.
    \end{aligned}
\end{equation}
This implies that $d_{0}\ket{i_0}=d_{1}\ket{i_1}$ and $e_{0}\ket{j_0}=-e_{1}\ket{j_1}$. Consequently, TP cannot distinguish between the states $d_{0}\ket{i_0}$ and $d_{1}\ket{i_1}$, and between the states $e_{0}\ket{j_0}$ and $-e_{1}\ket{j_1}$ unless TP sets all coefficients to zero: $d_0 = d_1 = e_0 = e_1 =0$. However, this contradicts the equations $\abs{d_0}^{2}+\abs{d_1}^{2}=1$ and $\abs{e_0}^{2}+\abs{e_1}^{2}=1$. Thus, TP cannot obtain any information about Alice's and Bob's secret key using his ancillary qubits. Therefore, there is no such $U_3$ operation to allow TP to obtain useful key information without being discovered if TP measured the qubits in the $X$ basis honestly.

In fact, TP may choose to measure the received qubits in bases other than the $X$ basis at this stage and disclose corresponding classical messages. Let's consider a scenario where, instead of measuring the received qubit from Bob in the $X$ basis, TP conducts a measurement in the $Z$ basis. If TP obtains the resultant state $\ket{0}$, he publishes 0; otherwise, he publishes 1. Referring to \hyperref[table 8]{Table 8}, to pass the detection, TP must set the incorrect items as zero vectors. Therefore, TP sets:

\begin{equation}
\label{35}
    d_{1}\ket{i_1} = e_{0}\ket{j_0} = 0.
\end{equation}
By doing so, state \Romannum{1} and state \Romannum{2} in \hyperref[table 8]{Table 8} become:

\begin{equation}\label{eq36}
    \begin{aligned}
        \ket{\psi_{\Romannum{1}}} = \ket{0}\ket{i_0}, \\
        \ket{\psi_{\Romannum{2}}} = \ket{1}\ket{j_1}.
    \end{aligned}
\end{equation}
Thus, TP can publish the correct measurement result and avoid detection. However, by fixing the condition in \hyperref[eq35]{equation (35)}, state \Romannum{3} and state \Romannum{4} in \hyperref[table 8]{Table 8} become:

\begin{equation}\label{eq37}
    \begin{aligned}
        \ket{\psi_{\Romannum{3}}} = \frac{1}{\sqrt{2}}\{\ket{0}\ket{i_0} + \ket{1}\ket{j_1}\}, \\
        \ket{\psi_{\Romannum{4}}} = \frac{1}{\sqrt{2}}\{\ket{0}\ket{i_0} - \ket{1}\ket{j_1}\}.
    \end{aligned}
\end{equation}

Conducting a $Z$ basis measurement, TP would obtain the state $\ket{0}\ket{i_0}$ or $\ket{1}\ket{j_1}$ with equal probability regardless of whether TP receives state \Romannum{3} or state \Romannum{4}. Consequently, TP would be unable to differentiate between case 3 and case 6 of \hyperref[table 8]{Table 8}, and therefore TP cannot measure his ancillary qubit later to determine Alice's and Bob's measurement outcomes in cases essential for establishing the raw key. Thus, no such $U_3$ operation exists that would allow TP to obtain useful key information without being detected if TP measures the qubits in the $Z$ basis. Similar analysis can be derived if TP measures the received qubits in other bases.

\subsubsection{Attacking strategy 2}\label{subsubsec3.3.2}

In this attacking strategy, TP consistently applies the three distinct $U_i$ operations on the same ancillary qubit and the intercepted qubit in each iteration. TP then retains the ancillary qubit in his quantum memory and measures it later on to extract useful information about Alice's and Bob's shared secret key.

Initially, TP prepares the state $\ket{+}$ and performs the $U_1$ operation on the state $\ket{+}$ and his ancillary qubit, initially in the state $\ket{e}$, resulting in:

\begin{equation}
\label{eq38}
    U_{1}(\ket{+}\otimes\ket{e}) = a_{0}\ket{0}\ket{f_0} + a_{1}\ket{1}\ket{f_1},
\end{equation}
where $\abs{a_0}^{2}+\abs{a_1}^{2}=1$, and $\{\ket{f_0}, \ket{f_1}\}$ are arbitrary quantum states, which do not necessarily need to be orthogonal, as TP won't use this basis as a measurement basis to obtain information about Alice's measurement outcomes. TP then sends the first qubit to Alice, who performs either of the operations: \textbf{(MH)} she measures the received qubit in the $Z$ basis, prepares a new qubit corresponding to the measurement result, then performs a Hadamard operation before sending it to Bob; and \textbf{(HM)} she performs a Hadamard operation on the received qubit, measures it in the $Z$ basis, prepares a new qubit corresponding to the measurement result, and sends it to Bob.

We know from \hyperref[table 1]{Table 1} that when Alice selects the \textbf{HM} operation, she expects to consistently obtain the state $\ket{0}$ upon measurement. However, observing the different measurement outcomes in \hyperref[table 6]{Table 6}, Alice has a probability of $\frac{1}{2}$ of obtaining the state $\ket{1}$, resulting in the protocol being aborted.

To prevent this scenario, TP needs to set the incorrect term, $a_{0}\ket{f_0}-a_{1}\ket{f_1}$, to be a zero vector. This condition results in $a_{0}\ket{f_0}=a_{1}\ket{f_1}$. Since the states $\ket{f_0}$ and $\ket{f_1}$ are not required to be orthogonal, unlike in the first attacking strategy, fulfilling this condition is feasible. TP can simply set $a_0 = a_1 = \frac{1}{\sqrt{2}}$, which satisfy the equation $\abs{a_0}^{2}+\abs{a_1}^{2}=1$. By doing so, \hyperref[eq38]{equation (38)} becomes:

\begin{equation}\label{eq39}
    U_{1}(\ket{+}\otimes\ket{e}) = \frac{1}{\sqrt{2}}(\ket{0}+\ket{1})\ket{f_0} = \ket{+}\ket{f_0},
\end{equation}
After Alice's operation, the state of the quantum system changes, as illustrated in \hyperref[table 9]{Table 9}.

\begin{table}
\caption{\label{table 9} The state of the quantum system (QS) after Alice's operations.}
\tiny
\resizebox{\textwidth}{!}{%
\begin{tabular}{ccccc}
\hline\hline
 Case & Alice's operation & \thead{\tiny State prepared \\ \tiny by Alice} & \thead{\tiny State of the QS after \\ \tiny Alice's operation} & \thead{\tiny Probability of this \\ \tiny case happening} \\
\hline
 1 & MH & $\ket{+}$ & $\ket{+}\ket{f_0}$ & $\cfrac{1}{4}$ \\
 2 & MH & $\ket{-}$ & $\ket{-}\ket{f_0}$ & $\cfrac{1}{4}$ \\
 3 & HM & $\ket{0}$ & $\ket{0}\ket{f_0}$  & $\cfrac{1}{2}$ \\

\hline\hline
\end{tabular}%
}

\end{table}

TP intercepts each qubit sent by Alice and performs the $U_2$ operation on the state of this qubit and his ancillary state. The $U_2$ operation is defined as follows:

\begin{equation}\label{eq40}
\begin{aligned}
U_{2}(\ket{+}\otimes\ket{f_0}) &= b_{0}\ket{0}\ket{g_0} + b_{1}\ket{1}\ket{g_1}, \\
U_{2}(\ket{-}\otimes\ket{f_0}) &= c_{0}\ket{0}\ket{h_0} + c_{1}\ket{1}\ket{h_1},
\end{aligned}
\end{equation}
where $\abs{b_0}^{2}+\abs{b_1}^{2}=1$, $\abs{c_0}^{2}+\abs{c_1}^{2}=1$, and $\{\ket{g_0}, \ket{g_1}\}$, $\{\ket{h_0}, \ket{h_1}\}$ do not need to be orthogonal since TP won't use those bases as a measurement basis to obtain information about Bob's measurement outcomes. By linearity, when TP conducts his attack on Alice's qubit in the case 3 of \hyperref[table 9]{Table 9}, $U_2$ is represented as:

\begin{equation}\label{eq41}
    U_{2}(\ket{0}\otimes\ket{f_0}) = \frac{1}{\sqrt{2}}\{\ket{0}(b_{0}\ket{g_0}+c_{0}\ket{h_0}) + \ket{1}(b_{1}\ket{g_1}+c_{1}\ket{h_1})\}.
\end{equation}

We notice that the states obtained are the same as in \hyperref[eq25]{equation (25)} and \hyperref[eq26]{equation (26)} in \hyperref[subsubsec3.3.1]{subsubsection 3.3.1}. After TP transmits the first qubit to Bob, and after Bob performs his operations, which are similar to those of Alice, we end up with exactly the same cases as in \hyperref[table 7]{Table 7}. Similar to the first attacking strategy, if we focus on the cases where Alice chooses the \textbf{HM} operation and Bob selects the \textbf{MH} operation, according to \hyperref[table 1]{Table 1} both Alice and Bob should consistently measure the state $\ket{0}$; otherwise, the protocol is aborted. Therefore, TP must avoid case 10 in \hyperref[table 7]{Table 7} to pass the public discussion. He then has to set the incorrect term $b_{1}\ket{g_1}+c_{1}\ket{h_1}$ to a zero vector as follows:

\begin{equation}\label{eq42}
         b_{1}\ket{g_1}+c_{1}\ket{h_1} = 0.
\end{equation}
Next, we consider the situation when Alice selects the \textbf{MH} operation and Bob chooses the \textbf{HM} operation. If we refer again to \hyperref[table 1]{Table 1}, the qubits involved in this scenario serve to establish the raw key due to the consistent correlation in Alice's and Bob's measurement outcomes. However, as shown in \hyperref[table 7]{Table 7}, TP's interference causes the measurement outcomes of Alice and Bob to be anti-correlated half of the time when they respectively choose the \textbf{MH} operation and \textbf{HM} operation, specifically in case 4 and case 7. To avoid this situation, TP has to set the terms $b_{0}\ket{g_0}-b_{1}\ket{g_1}$ and $c_{0}\ket{h_0}+c_{1}\ket{h_1}$ to zero vectors as follows:

\begin{equation}\label{eq43}
   \begin{aligned}
    b_{0}\ket{g_0} - b_{1}\ket{g_1} &= 0, \\
    c_{0}\ket{h_0} + c_{1}\ket{h_1} &= 0.
   \end{aligned}
\end{equation}
Consequently, by combining the two conditions in \hyperref[eq42]{equation (42)} and \hyperref[eq43]{equation (43)}, we will end up with exactly the same condition as with the first attacking strategy that TP must fulfill in order to avoid disrupting the original system and pass the public discussion:
\begin{equation}\label{eq44}
    b_{0}\ket{g_0}=b_{1}\ket{g_1}=c_{0}\ket{h_0}=-c_{1}\ket{h_1}.
\end{equation}

Since the states $\{\ket{g_0}, \ket{g_1}\}$, and $\{\ket{h_0}, \ket{h_1}\}$ are not required to be distinguishable, this condition can be fulfilled by setting the coefficients to be equal:
\begin{equation}\label{eq45}
    b_0 = b_1 = c_0 = -c_1 = \frac{1}{\sqrt{2}}
\end{equation}
This indeed satisfies the equations $\abs{b_0}^{2}+\abs{b_1}^{2}=1$, and $\abs{c_0}^{2}+\abs{c_1}^{2}=1$. Thus, \hyperref[eq40]{equation (40)} can be reduced to:

\begin{equation}\label{eq46}
\begin{aligned}
U_{2}(\ket{+}\otimes\ket{f_0}) &= \frac{1}{\sqrt{2}}(\ket{0}+\ket{1})\ket{g_0} = \ket{+}\ket{g_0}, \\
U_{2}(\ket{-}\otimes\ket{f_0}) &= \frac{1}{\sqrt{2}}(\ket{0}-\ket{1})\ket{g_0} = \ket{-}\ket{g_0},
\end{aligned}
\end{equation}
By linearity:

\begin{equation}\label{eq47}
    U_{2}(\ket{0}\otimes\ket{f_0}) = \ket{0}\ket{g_0}
\end{equation}
Following Bob's operation, the state of the quantum system changes, as illustrated in \hyperref[table 10]{Table 10}.

\begin{table}
\caption{\label{table 10} The state of the quantum system (QS) after Bob's operations.}
\resizebox{\textwidth}{!}{%
\begin{tabular}{ccccccc}
\hline\hline
 Case & Alice's operation & \thead{State prepared \\ by Alice} & Bob's operation & \thead{State prepared \\ by Bob} & State of the QS & \thead{Probability of this \\ case happening} \\
\hline
 1 & MH & $\ket{+}$ & MH & $\ket{+}$ & State \Romannum{1} & $\cfrac{1}{8} \abs{b_{0}}^2$ \\
 2 & MH & $\ket{+}$ & MH & $\ket{-}$ & State \Romannum{2} & $\cfrac{1}{8} \abs{b_{1}}^2$ \\
 3 & MH & $\ket{+}$ & HM & $\ket{0}$ & State \Romannum{3} & $\cfrac{1}{16}$ \\

 4 & MH & $\ket{-}$ & MH & $\ket{+}$ & State \Romannum{1} & $\cfrac{1}{8} \abs{c_{0}}^2$ \\
 5 & MH & $\ket{-}$ & MH & $\ket{-}$ & State \Romannum{2} & $\cfrac{1}{8} \abs{c_{1}}^2$ \\
 6 & MH & $\ket{-}$ & HM & $\ket{1}$ & State \Romannum{4} & $\cfrac{1}{16}$ \\
 
 7 & HM & $\ket{0}$ & MH & $\ket{+}$ & State \Romannum{1} & $\cfrac{1}{8}(\abs{b_0}^{2}+\abs{c_0}^{2})$ \\

 8 & HM & $\ket{0}$ & HM & $\ket{0}$ & State \Romannum{3} & $\cfrac{1}{8}$ \\
 9 & HM & $\ket{0}$ & HM & $\ket{1}$ & State \Romannum{4} & $\cfrac{1}{8}$ \\

\hline\hline
\end{tabular}%
}
\footnotesize
State \Romannum{1}: $\ket{+}\ket{g_0}$ \\
State \Romannum{2}: $\ket{-}\ket{g_0}$ \\
State \Romannum{3}: $\ket{0}\ket{g_0}$ \\
State \Romannum{4}: $\ket{1}\ket{g_0}$ \\

\end{table}

After Bob's operations, he forwards the qubit to TP, who then executes the $U_3$ operation on the received qubit and his ancillary qubit. This operation is defined as follows:
\begin{equation}\label{eq48}
    \begin{aligned}
         U_{3}(\ket{+}\otimes\ket{g_0}) = d_{0}\ket{0}\ket{i_0} + d_{1}\ket{1}\ket{i_1}, \\
         U_{3}(\ket{-}\otimes\ket{g_0}) = e_{0}\ket{0}\ket{j_0} + e_{1}\ket{1}\ket{j_1},
    \end{aligned}
\end{equation}
where $\abs{d_0}^{2}+\abs{d_1}^{2}=1$, $\abs{e_0}^{2}+\abs{e_1}^{2}=1$, and $\{\ket{i_0}, \ket{i_1}\}$, $\{\ket{j_0}, \ket{j_1}\}$ are orthogonal bases that TP can distinguish. This results in:
\begin{equation}
\label{eq50}
    \begin{aligned}
         U_{3}(\ket{0}\otimes\ket{g_0}) = \frac{1}{\sqrt{2}}\{\ket{0}(d_{0}\ket{i_0}+e_{0}\ket{j_0}) + \ket{1}(d_{1}\ket{i_1}+e_{1}\ket{j_1})\}, \\
         U_{3}(\ket{1}\otimes\ket{g_0}) = \frac{1}{\sqrt{2}}\{\ket{0}(d_{0}\ket{i_0}-e_{0}\ket{j_0}) + \ket{1}(d_{1}\ket{i_1}-e_{1}\ket{j_1})\}.
    \end{aligned}
\end{equation}
These states are identical to those in \hyperref[eq31]{equation (31)} and \hyperref[eq32]{equation (32)} in the first attacking strategy, indicating that the rest of this analysis of TP's attack will be the same as when we analyze TP's attack on the quantum channel between Bob and TP in the first attacking strategy.

Following the $U_3$ operation, the state of the quantum system changes, as illustrated in \hyperref[table 8]{Table 8}. Subsequently, TP can measure the received qubit in any single-qubit basis of his choice. Here, we will analyze two scenarios. In the first scenario, TP selects the $X$ basis according to the protocol. If he obtains the measurement result $\ket{+}$, he publishes 0; otherwise, he publishes 1. To avoid detection, TP must obtain the same measurement result as the state sent by Bob when Bob chooses the \textbf{MH} operation. States \Romannum{1} and \Romannum{2} in \hyperref[table 8]{Table 8} are rewritten as:
\begin{equation}\label{eq50}
    \begin{aligned}
         \ket{\psi_{\Romannum{1}}} = \frac{1}{\sqrt{2}}\{\ket{+}(d_{0}\ket{i_0}+d_{1}\ket{i_1}) + \ket{-}(d_{0}\ket{i_0}-d_{1}\ket{i_1})\}, \\
         \ket{\psi_{\Romannum{2}}} = \frac{1}{\sqrt{2}}\{\ket{+}(e_{0}\ket{j_0}+e_{1}\ket{j_1}) + \ket{-}(e_{0}\ket{j_0}-e_{1}\ket{j_1})\}.
    \end{aligned}
\end{equation}
For TP to pass detection, he must set the incorrect items to zero vectors:

\begin{equation}\label{eq51}
    \begin{aligned}
        d_{0}\ket{i_0}-d_{1}\ket{i_1} = 0, \\
        e_{0}\ket{j_0}+e_{1}\ket{j_1} = 0.
    \end{aligned}
\end{equation}
This implies that $d_{0}\ket{i_0}=d_{1}\ket{i_1}$ and $e_{0}\ket{j_0}=-e_{1}\ket{j_1}$. Consequently, TP cannot distinguish between the states $d_{0}\ket{i_0}$ and $d_{1}\ket{i_1}$, and between the states $e_{0}\ket{j_0}$ and $-e_{1}\ket{j_1}$ unless TP sets all coefficients to zero: $d_0 = d_1 = e_0 = e_1 =0$. However, this contradicts the equations $\abs{d_0}^{2}+\abs{d_1}^{2}=1$ and $\abs{e_0}^{2}+\abs{e_1}^{2}=1$. Thus, TP cannot obtain any information about Alice's and Bob's secret key by measuring his ancillary qubits.

In the second scenario, TP conducts a measurement in the $Z$ basis. If TP obtains the resultant state $\ket{0}$, he publishes 0; otherwise, he publishes 1. Referring to \hyperref[table 8]{Table 8}, to avoid detection, TP must set the incorrect items to zero vectors. Therefore, TP sets:

\begin{equation}
\label{eq52}
    d_{1}\ket{i_1} = e_{0}\ket{j_1} = 0.
\end{equation}
By doing so, state \Romannum{1} and state \Romannum{2} in \hyperref[table 8]{Table 8} become:

\begin{equation}\label{eq52}
    \begin{aligned}
        \ket{\psi_{\Romannum{1}}} = \ket{0}\ket{i_0}, \\
        \ket{\psi_{\Romannum{2}}} = \ket{1}\ket{j_1}.
    \end{aligned}
\end{equation}
Thus, TP can publish the correct measurement result and avoid detection. However, by fixing the condition in \hyperref[eq52]{equation (52)}, states \Romannum{3} and \Romannum{4} in \hyperref[table 8]{Table 8} become:

\begin{equation}\label{eq53}
    \begin{aligned}
        \ket{\psi_{\Romannum{3}}} = \frac{1}{\sqrt{2}}\{\ket{0}\ket{i_0} + \ket{1}\ket{j_1}\}, \\
        \ket{\psi_{\Romannum{4}}} = \frac{1}{\sqrt{2}}\{\ket{0}\ket{i_0} + \ket{1}\ket{j_1}\}.
    \end{aligned}
\end{equation}
Performing a measurement in the $Z$ basis, TP would equally likely obtain either the state $\ket{0}\ket{i_0}$ or $\ket{1}\ket{j_1}$, regardless of whether TP receives state \Romannum{3} or state \Romannum{4}. Consequently, TP would be unable to differentiate between case 3 and case 6 of \hyperref[table 8]{Table 8}, and thus cannot discern Alice's and Bob's measurement outcomes in the cases that serve to establish the raw key.

As a result, we have demonstrated that there are no $U_1$, $U_2$, and $U_3$ operations that would allow TP to gather useful key information without having a non-zero probability of being detected, regardless of whether TP chooses to measure the received qubits in the $X$ basis or in the $Z$ basis. A similar analysis can be conducted if TP measures the received qubits in other bases.

\section{Comparisons}\label{sec4}

In this section, we undertake a comparison between our protocol and several analogous MSQKD protocols, including Krawec's protocol \cite{Krawec2015}, Liu et al.'s protocol \cite{Liu2018}, Lin et al.'s protocol \cite{Lin2019}, Tsai's and Yang's protocol \cite{Tsai2021}, Chen et al.'s protocol \cite{Chen2021}, and Ye et al.'s protocol \cite{Ye2022}. The comparisons are drawn from four perspectives: quantum resources, the TP's capabilities, operations permitted for classical participants, and qubit efficiency, defined as the ratio of the number of raw key bits to the total qubits utilized. The comparison results are presented in \hyperref[table 11]{Table 11}.

If we first compare the abilities of the classical participants, we see that only in our protocol and the one proposed by Tsai and Yang do classical participants possess the ability to perform the Hadamard operation. Although Tsai's and Yang's protocol exempt the classical participants from generating qubits in the $Z$ basis, the quantum overhead of the TP is heavier since the preparation of Bell states is more difficult and expensive than the generation of single qubits in practice. On the other hand, in Liu et al.'s protocol, classical participants are not required to perform quantum measurements in the $Z$ basis. However, they need quantum delay lines to store the qubits for a certain period of time for reordering them. 

Regarding quantum resources, besides our protocol and those proposed by Lin et al. and Chen et al., the TP necessitates the utilization of Bell states as quantum resources to aid in the distribution of secret keys between Alice and Bob. Examining the TP's capabilities, we observe that apart from our protocol and the one proposed by Chen et al., all protocols mandate the TP to at least prepare or measure qubits in the Bell basis. In contrast, our protocol, along with Chen et al.'s protocol, solely requires the TP to prepare the $\ket{+}$ state and conduct measurements in the $X$ basis $\{\ket{+}, \ket{-}\}$. This reduction in the quantum resources and the quantum workload of the TP enhances the feasibility and practicality of the protocol in real-world scenarios. 

In terms of qubit efficiency, our proposed MSQKD protocol outperforms Krawec’s, Ye et al.'s, and Lin et al.’s protocols, yet matches the efficiency of Liu et al.’s, Chen et al.'s, and Tsai's and Yang's protocols. Although the efficiency of our proposed protocol can be enhanced by setting the probability of Alice choosing the \textbf{MH} operation and Bob selecting the \textbf{HM} operation above $\frac{1}{2}$.

Furthermore, our protocol exhibits significant similarities with Chen et al.'s protocol, differing primarily in the operations performed by the classical participants. In our protocol, participants must choose between the \textbf{MH} operation and the \textbf{HM} operation at each round, whereas in Chen et al.'s scheme, participants choose between the reflect operation and the measure and resend operation. This difference leads to an increase in the number of cases available for checking TP's honesty. In both protocols, depending on Alice's and Bob's operations, four cases are possible. However, in Chen et al.'s protocol, only the case where both Alice and Bob choose the reflect operation can be used to check the TP's honesty and detect potential eavesdroppers, whereas in our protocol, three out of the four cases can be utilized for this purpose. This results in higher probabilities of detecting TP's various attacks.

\begin{table}
\caption{\label{table 11} Comparisons of the MKSQD protocol.}
\resizebox{\textwidth}{!}{%
\begin{tabular}{ccccc}
\hline\hline
 Protocols & Classical participant's ability & Quantum resources & TP’s quantum ability & Qubit efficiency \\
\hline
 Krawec's\cite{Krawec2015}& \makecell{(1) Generate $\ket{0}$ or $\ket{1}$ \\ (2) Measure in $Z$ \\ (3) Reflect} & Bell states & \makecell{(1) Prepare Bell states \\ (2) Perform Bell measurement} & $\cfrac{1}{24}$ \\
 \hline

 Liu et al.'s\cite{Liu2018} & \makecell{(1) Generate $\ket{0}$ or $\ket{1}$ \\ (2) Reorder \\ (3) Reflect} & \makecell{(1) Bell states \\ (2) Single qubits} & \makecell{(1) Prepare Bell states \\ (2) Perform Bell measurement} &  $\cfrac{1}{8}$ \\
 \hline

 Lin et al.'s\cite{Lin2019} & \makecell{(1) Generate $\ket{0}$ or $\ket{1}$ \\ (2) Measure in $Z$ \\ (3) Reflect} & Single qubits & \makecell{(1) Prepare $\ket{+}$ \\ (2) Perform Bell measurement} & $\cfrac{1}{24}$ \\
 \hline

 Tsai's and Yang's\cite{Tsai2021} & \makecell{(1) Measure in $Z$ \\ (2) Perform $H$} & Bell states & Prepare Bell states & $\cfrac{1}{8}$ \\
 \hline

 Chen et al.'s\cite{Chen2021} & \makecell{(1) Generate $\ket{0}$ or $\ket{1}$ \\ (2) Measure in $Z$ \\ (3) Reflect} & Single qubits & \makecell{(1) Prepare $\ket{+}$ \\ (2) Measure in the $X$ basis} & $\cfrac{1}{8}$ \\
 \hline

 Ye et al.'s\cite{Ye2022} & \makecell{(1) Generate $\ket{0}$ or $\ket{1}$ \\ (2) Measure in $Z$ \\ (3) Reflect} & Bell states & \makecell{(1) Prepare Bell states \\ (2) Perform Bell measurement} & $\cfrac{1}{12}$ \\
 \hline

 Our protocol & \makecell{(1) Generate $\ket{0}$ or $\ket{1}$ \\ (2) Measure in $Z$ \\ (3) Perform $H$} & Single qubits & \makecell{(1) Prepare $\ket{+}$ \\ (2) Measure in the $X$ basis} & $\cfrac{1}{8}$ \\
 
\hline\hline
\end{tabular}%
}
\end{table}

\section{Conclusion}\label{sec5}

In this study, we have proposed an efficient MSQKD protocol designed to enable two "classical" participants to securely establish a shared secret key with the assistance of an untrusted TP. Notably, our protocol imposes more limited quantum capabilities on the TP compared to existing similar MSQKD protocols, while maintaining a respectable qubit efficiency, enhancing its practicality in real-world applications. We have also shown that the proposed protocol can withstand various well-known attacks, including the measurement attack, the faked states attack, and the collective attack. In these scenarios, an attacker is unable to extract useful information about Alice's and Bob's shared secret key without detection. However, our security analysis was conducted under ideal conditions, assuming noise-free and non-lossy quantum channels. Further research to extend this analysis to noisy environments warrants further investigation.

\section*{Acknowledgment}

This research did not receive any specific grant from funding agencies in the public, commercial, or not-for-profit sectors.

\bibliographystyle{plainnat}
 \bibliography{msqkd_ref}

\end{document}